\documentclass[useAMS,usenatbib,usegraphicx]{mn2e}

			\usepackage{amsmath}
			\usepackage{amssymb}
		
			\usepackage{times}

			\usepackage{natbib}
			\usepackage{url}	% to display ULRs correctly
		
			\newcommand{\chandra}{\textit{Chandra}}
			\newcommand{\xmm}{XMM-\textit{Newton}}
			\newcommand{\erosita}{SRG/eROSITA}
			
			\newcommand{\PoSp}{power spectrum}
			\newcommand{\PoSpa}{power spectra}
			
			\newcommand{\mPoSp}{mosaic \PoSp{}}
			\newcommand{\sPoSp}{stacked \PoSp{}}
			\newcommand{\cPoSp}{combined \PoSp{}}
			\newcommand{\cxbPS}{CXB \PoSp{}}
			\newcommand{\lssPS}{LSS \PoSp{}}

			\newcommand{\OHT}{one-halo term}

			\newcommand{\PhSN}{photon shot noise}
			\newcommand{\PoSoSN}{point-source shot noise} 
			
			\newcommand{\ePart}{\ensuremath{9.5-12.0\,\mathrm{keV}}}
			\newcommand{\eSoft}{\ensuremath{0.5-2.0\,\mathrm{keV}}}
			\newcommand{\eHard}{\ensuremath{2.0-7.0\,\mathrm{keV}}}
			
			\newcommand{\eGal}{\ensuremath{0.5-0.7\,\mathrm{keV}}}

			\newcommand{\eBol}{\ensuremath{0.1-100\,\mathrm{keV}}}
			
			\newcommand{\ePartB}{\ePart{} band}
			\newcommand{\eSoftB}{\eSoft{} band}

			\newcommand{\Cref}[1]{Chapter~\ref{#1}}
			\newcommand{\Sref}[1]{Section~\ref{#1}}
			\newcommand{\Fref}[1]{Figure~\ref{f:#1}}
			\newcommand{\Tref}[1]{Table~\ref{#1}}
			\newcommand{\Eref}[1]{Eq.~\ref{#1}}
			\newcommand{\Eqref}[1]{Eq.~\eqref{#1}}
			\newcommand{\Aref}[1]{Appendix~\ref{#1}}

			\renewcommand{\d}{\ensuremath{\mathrm{d}}} 
			
			\newcommand{\InvArcSec}{\ensuremath{\,\mathrm{arcsec^{-1}}}}
			
			\newcommand{\LogNLogS}{\ensuremath{\log N - \log S}}
			
			\newcommand{\NH}{\ensuremath{N_\mathrm{H}}}

			\newcommand{\APEC}{\texttt{APEC}}
			\newcommand{\xspec}{\texttt{XSPEC}}
			
			\newcommand{\map}[1]{\ensuremath{\textit{\textbf{#1}}}}
			\newcommand{\msc}[1]{\ensuremath{\mathbb{#1}}}

			\newcommand{\GCG}{clusters of galaxies}
			\newcommand{\rES}{resolved extended sources}
			\newcommand{\rPS}{resolved point sources}

			\newcommand{\acisi}{\mbox{ACIS-I}}
	
			\newcommand{\PapIII}{Paper~III}
			
			\newcommand{\FigRoot}{p67_MSC_h_TalkPlots_Pix32_FOV092_}

\begin{document}

	%===========================================================
	% some makros
	%===========================================================
		\defcitealias{Kolodzig2016}{Paper~I}
		\newcommand{\Kol}{\citetalias{Kolodzig2016}}
	
		\defcitealias{Kenter2005}{K05}
		\newcommand{\Ken}{\citetalias{Kenter2005}}
	
	%======================================================================================================================
	% Title + Abstract
	%======================================================================================================================
		\title[Studying the ICM in \GCG{}]{Studying the ICM in \GCG{} via surface brightness fluctuations of the cosmic X-ray background}
		
		\author[A. Kolodzig et al.]{
			Alexander~Kolodzig$^{1}$,
			Marat~Gilfanov$^{2,3}$,
			Gert~H\"{u}tsi$^{4}$,
			Rashid~Sunyaev$^{2,3}$ \\
			$^1$Kavli Institute for Astronomy and Astrophysics (KIAA), Peking University, 100871 Beijing, China --  KIAA fellow, \url{alex@kolodzig.eu} \\
			$^2$Max-Planck-Institut f\"{u}r Astrophysik (MPA), Karl-Schwarzschild-Str. 1, D-85741 Garching, Germany \\
			$^3$Space Research Institute (IKI), Russian Academy of Sciences, Profsoyuznaya ul. 84/32, Moscow, 117997 Russia \\
			$^4$Tartu Observatory, T\~oravere 61602, Estonia \\
		}
		
		\date{Accepted 2O!7 Xxx XX. Received 2O!7 Xxx XX; in original form 2O!7 Xxx XX}
		%!!! The proper receipt and acceptance dates of your manuscript will be set by the editors and inserted by the publisher.
		
		\pagerange{\pageref{firstpage}--\pageref{lastpage}} \pubyear{2O!7}
		
		\maketitle
		\label{firstpage}

		\begin{abstract} %
			We study the surface brightness fluctuations of the cosmic X-ray background (CXB) using Chandra data of XBOOTES. 
			After masking out resolved sources we compute the power spectrum of fluctuations of the unresolved CXB for angular scales from $\approx2\arcsec$ to $\approx3^\circ$.
			The non-trivial large-scale structure (LSS) signal dominates over the shot noise of unresolved point sources at all scales above $\sim1\arcmin$ and is produced mainly by the intracluster medium (ICM) of unresolved clusters and groups of galaxies, as shown in our previous publication.

			The shot-noise-subtracted power spectrum of CXB fluctuations has a power-law shape with the slope of $\Gamma=0.96\pm0.06$. 
			Its energy spectrum is well described by the redshifted emission spectrum of optically-thin plasma with the best-fit temperature of $T\approx1.3$~keV and the best-fit redshift of $z\approx0.40$. 
			They are in good agreement with theoretical expectations based on the X-ray luminosity function and scaling relations of clusters. 
			From these values we estimate the typical mass and luminosity of the objects responsible for CXB fluctuations, $M_{500}\sim10^{13.6}\,\mathrm{M_{\sun}}\,h^{-1}$ and $L_{0.5-2.0\,{\rm keV}}\sim10^{42.5}\,\mathrm{erg\,s^{-1}}$. 
			On the other hand, the flux-weighted mean temperature and redshift of resolved clusters are $T\approx2.4$~keV and $z\approx0.23$, confirming that fluctuations of unresolved CXB are caused by cooler (i.e. less massive) and more distant clusters, as expected. 
			We show that the power spectrum shape is sensitive to the ICM structure all the way to the outskirts, out to $\sim{\rm few}\times R_{500}$. 
			We also look for possible contribution of the warm-hot intergalactic medium (WHIM) to the observed CXB fluctuations.

			Our results underline the significant diagnostics potential of the CXB fluctuation analysis in studying the ICM structure in clusters.
		\end{abstract}
		
		\begin{keywords}
			-- large-scale structure of Universe
			-- X-rays: diffuse background
			-- galaxies: clusters: intracluster medium
			-- galaxies: clusters: general
			-- galaxies: groups: general
			-- galaxies: active
		\end{keywords}

	%======================================================================================================================
	% Content
	%======================================================================================================================

% 	
	%---------------------------------------------------------------------------------------------------------------------- '
	\section{Introduction} \label{s:intro}
	%----------------------------------------------------------------------------------------------------------------------
		Since the discovery of the cosmic X-ray background (CXB) more than half a century ago \citep{Giacconi1962},
		analyzing its surface brightness fluctuations via angular correlation studies has been a powerful tool in understanding the origin of the CXB \citep[e.g.][]{Scheuer1974,Hamilton1987,Shafer1983,Barcons1988,Soltan1994,Vikhlinin1995b,Miyaji2002}.
		Such CXB fluctuation analyses have suggested very early on that the CXB is dominated by extragalactic discrete sources, with Active Galactic Nuclei (AGN) leading the way, and that their redshift distribution is similar to optical QSOs but with somewhat higher clustering strength.
		These results were confirmed during  the last $\sim$two decades with the resolved  X-ray sources  through  their source  counts with very deep pencil beam surveys \citep[e.g.][]{Brandt2005,Alexander2013,Brandt2015,Lehmer2012,Luo2017}
		and large-scale structure (LSS) studies with much wider but  shallower surveys  \citep[see reviews of][]{Cappelluti2012,Krumpe2013}.
		
		The CXB is an ideal laboratory for studying growth and co-evolution of supermassive black holes (SMBH) with their dark matter halo (DMH) up to high redshift \citep[$z\sim5$, e.g.][]{Hasinger2005,Gilli2007,Aird2010,Ueda2014,Miyaji2015},
		which is a keystone in understanding galaxy evolution over cosmic time \citep[e.g.][]{Hopkins2006,Hickox2009,Alexander2012,Heckman2014}.
		Thanks to the high AGN number density  and efficiency of their detection in X-ray surveys, it will soon  become possible to use large samples of X-ray-selected AGN as a cosmological probe via baryon acoustic oscillation (BAO) measurements \citep{Kolodzig2013,Huetsi2013}. 
		In particular,  this should become achievable with the  $\sim3$~million X-ray selected AGN to be detected in the upcoming \erosita{} all-sky survey \citep[eRASS,][]{eROSITA,eROSITA.SB,Kolodzig2012}.
		
		The field of CXB fluctuation analysis is currently undergoing a renaissance,
		thanks to the availability of  X-ray surveys of various area and depth with superb angular resolution conducted by \chandra{} and \xmm{}  X-ray observatories \citep[see reviews of][]{Brandt2005,Brandt2015}.
		The first studies of this kind focused on the fluctuations of  unresolved CXB in \chandra{}'s  deep surveys.
		Since such surveys have a very small sky coverage ($\lesssim0.1\,\mathrm{deg^2}$), the analyses were limited to small angular scales below \chandra{} \acisi{}'s FOV ($\lesssim17\arcmin$).
		The study of \citet{Cappelluti2012b} used the $\sim4$~Ms \chandra{} Deep Field-South Survey \citep[CDF-S, $\sim0.02$~deg$^2$,][]{Xue2011}, and associated the detected fluctuation signal with a combined contribution of unresolved AGN, galaxies and the intergalactic medium.
		The subsequent study by \citet{Cappelluti2013,Helgason2014} used the $\sim0.6$~Ms \chandra{} AEGIS-XD survey \citep[$\sim0.1$~deg$^2$,][]{Goulding2012} and concluded that the fluctuation signal is dominated by
		the shot noise of unresolved AGN. 
		 They also detected a significant excess at the angular scales of $\sim2\arcmin-3\arcmin$.
		Its origin, however, was not further investigated as it was not relevant to the main focus of their work, which was a possible clustering signal of very high redshift ($z>5$) AGN via a cross-correlation analysis with the cosmic near-infrared (NIR) background \citep[also see e.g.][]{Yue2013,Yue2016,Helgason2016,Mitchell2016}.
		
		The most recent study of \citet[hereafter \Kol{}]{Kolodzig2016} used XBOOTES \citep[hereafter \Ken]{Murray2005,Kenter2005},
		the currently largest available continuous \chandra{} \acisi{} survey.
		It covers a surface area of $\sim9\,\mathrm{deg^2}$ of the B\"ootes field of the NOAO Deep Wide-Field Survey \citep[NDWFS,][]{Jannuzi2004}
		and has a depth of $\sim5$~ks.
		Based on this data, \Kol{} conducted the most accurate measurement to date of the \PoSp{} of fluctuations of the unresolved CXB.
		In their work they focused on angular scales below $\lesssim17\arcmin$ and could show that for angular scales below $\sim1\arcmin$ the \PoSp{} is consistent with the shot noise of unresolved AGN without any detectable contribution from their \OHT{}.
		However, at larger angular scales they detected a significant {power} above the AGN shot noise,
		which they associated with the  intracluster medium (ICM) of unresolved clusters and groups%
			\footnote{
			For simplicity, we will use in the following  the term '\GCG{}' to address both clusters and groups of galaxies.  
			Note that there is no formal sharp  separation between clusters and groups of galaxies.
			The smallest groups have a mass of the order of $M_{500} \sim 10^{12}\,\mathrm{M_{\sun}}$,
			while the largest clusters can reach of the order of $M_{500} \sim 10^{15}\,\mathrm{M_{\sun}}$  \citep[e.g.][]{Kravtsov2012}.
			}
		of galaxies based on several observational and theoretical evidences.
		
		The ICM has a typical  temperature in the keV-regime and emits X-rays through emission mechanisms of optically thin plasma.
		Its X-ray surface brightness is determined by the gas  temperature and density distributions,
		which are tightly correlated with the density profile of its underlying DMH \citep[e.g.][]{Komatsu2001}.
		These dependencies are exploited by creating scaling relations between ICM observables and the DMH mass in order to measure the spatial density of DMHs as a function of their mass and redshift, alias the halo mass function 
		\citep[e.g.][]{Vikhlinin2006,Sun2009,Ettori2013,Giodini2013}.
		The halo mass function {is an important probe for the key cosmological parameters of the Universe,
		which makes its measurement one of the main science drivers, along with the studies of the AGN and quasar populations, } for very large X-ray surveys, such as XXL \citep[$\sim25\times2$~deg$^2$,][]{XXL2016} and eRASS.
		However, {accurate calibration of  the scaling relations is a challenging task}, 
		because, among others, the most common assumptions of a hydrostatic equilibrium and spherical symmetry of the ICM are {significant} simplifications \citep[e.g.][]{Giodini2013}.
		The ICM has a rich structure, which is the result of a complex interplay between gravity-induced dynamics
		and non-gravitational processes (e.g. AGN feedback, radiative cooling, star formation, and galactic winds).
		This makes studies of the ICM structure {of primary importance  not only for understanding the formation and evolution of galaxies, but also for the  cosmological measurements}  \citep[see reviews of][]{Rosati2002,Kravtsov2012}.
				
		% AIM of Paper
		Studying the ICM structure of a very large sample of resolved \GCG{} via X-ray surface-brightness profile measurement is observationally very expensive \citep[e.g.][]{Eckert2012,Eckert2016a,XXL2016}.
		Based on the results of \Kol{}, we investigate in this work  the potential of using CXB fluctuation analysis for ICM, 
		which may lead to important improvements in our understanding of {its} structure, especially at the outskirts of \GCG{}. 
		
		We will primarily focus on the CXB fluctuations at large angular scales up to the XBOOTES limit of $\sim3^\circ$,
		which have not been studied in \Kol{}.
		To this end, we construct a mosaic image of XBOOTES to compute the \emph{\mPoSp{}}, as opposite to the \emph{\sPoSp{}} of individual XBOOTES observations computed in \Kol{}. In our analysis, we will obtain the power spectra of resolved \GCG{} along with the unresolved part of the CXB. 
		We will also obtain  the energy spectra of fluctuations and compare them with theoretical expectations for both unresolved and resolved \GCG{}.
		We compare the measured power spectra with theoretical predictions of the clustering signal of \GCG{} in \PapIII{} (in prep.).

		% wHIM
		The warm-hot intergalactic medium (WHIM)  is expected  to account for almost a half of the baryonic matter in the  Universe.
		Its hottest fraction located in the unvirialized outskirts of \GCG{} and connecting filaments  is shock-heated  to the sub-keV temperatures \citep[e.g.][]{Dave2001,Bregman2007},
		and can make a non-negligible contribution to the surface brightness of  unresolved CXB and its fluctuations, 
		as cosmological hydrodynamical simulations and very deep ($>100$~ks) X-ray observations {seem to suggest} \citep[e.g.][]{Hickox2007,Werner2008,Galeazzi2009,Roncarelli2006,Roncarelli2012,Ursino2011,Ursino2014,Nevalainen2015,Eckert2015}.
		Given its very faint and diffuse nature, it is very difficult { to be observed directly.}
		Therefore, various other methods such as CXB fluctuation analysis have been proposed in order to study its properties \citep[e.g.][]{Kaastra2013}.
		We investigate in this work whether it is possible to detect the CXB fluctuations {due to  WHIM } with an XBOOTES-like survey.
		
		% structure
		This paper is organized as following:
		In \Sref{s:DataProc} we explain our data processing procedure, 
		in \Sref{s:Fluc} we present the \PoSp{} of the CXB surface brightness fluctuations and we study its properties in \Sref{s:LSS_PL}. 
		Our results are summarized in \Sref{s:sum}.
		In  Appendixes  we present results of tests for various systematic effects and investigate the impact of the instrumental background on our measurements.
		% Cosmology
		{For consistency} we assume the same {flat $\Lambda$CDM} cosmology as in \Kol{}:
		{$H_0 = 70\,\mathrm{km\,s^{-1}\,Mpc^{-1}}$ ($h=0.70$),
		$\Omega_\mathrm{m} = 0.30$ ($\Omega_\Lambda = 0.70$),
		$\Omega_\mathrm{b} = 0.05$, %$\Omega_\mathrm{k} = 0.00$,
		$\sigma_8=0.8$}.

	%----------------------------------------------------------------------------------------------------------------------
	\section{Data preparation and processing} \label{s:DataProc}
	%----------------------------------------------------------------------------------------------------------------------
		As in \Kol{}, we are using in this work the $\sim5$~ks deep, $\sim9$~deg$^2$ large \chandra{} \acisi{} survey XBOOTES \citep[\Ken]{Murray2005}.
		We  adopt the data preparation and processing procedures  from \Kol{} (section~2) with a few important changes and additional steps, which we describe below.
		These changes are necessary because {we are computing the \PoSp{} of the $\sim 3^\circ\times 3^\circ$  mosaic image using all observations of the of XBOOTES field as opposite to the \sPoSp{} of individual \chandra{} observations ($\sim 17\arcmin\times 17\arcmin$ in size), considered in the \Kol{}. Also, we are now using the entire energy band of \acisi{} ($0.5-10.0\,\mathrm{keV}$).  
		For consistency, we will also recompute the \sPoSp{} of individual observations, which will be used to characterize the high frequency part of the final \PoSp{}. 
		In constructing the mosaic image we will use all \chandra{} observations of the XBOOTES field, whereas in \Kol{} we excluded several observations.
		These changes do not have any significant impact on the results presented in \Kol{}, as it is demonstrated in \Aref{a:Data} where  we compare the stacked power spectra obtained in this work and those from \Kol{}.}  
				
		{Unless}  stated otherwise we use the spectral model of the unresolved CXB from \Kol{} (section~3) to convert between physical and instrumental units, and assume for the Galactic absorption a hydrogen column density of $\NH=10^{20}\,\mathrm{cm^{-2}}$ \citep[\Ken]{Kalberla2005} and  the metallicity of $0.3$ of the solar value \citep{Anders1989}.

	\subsection{Changes to \Kol{}} \label{s:Change}
	%-----------------------------------------------------------
		
		\subsubsection{Field selection} \label{s:F_sel}
		In this work we are using  all 126 individual, contiguous \chandra{} \acisi{} observations of XBOOTES, while in \Kol{} eight of them (3601, 3607, 3617, 3625, 3641, 3657, 4224 \& 4228) were excluded for various reasons (see section~2 of \Kol{}). 
				
		\subsubsection{Exposure map and FOV mask} \label{s:ETM_MSK}
		The exposure map $\map{E}$ (seconds) and the mask $\map{M}$ are now computed for each energy band individually.
 		This becomes necessary because we are now also studying fluctuations at higher energy bands, where
		the effective area\footnote{\label{acisi_area}\url{http://cxc.harvard.edu/proposer/POG/html/chap6.html#fig:acis_effarea_lin}}
		and vignetting\footnote{\label{acisi_vig}\url{http://cxc.harvard.edu/proposer/POG/html/chap6.html#tth_fIg6.6}} of \acisi{} 
 		can be quite different in comparison to the \eSoftB{}.
		{As before,} for the \ePartB{} we are adopting the mask and the average exposure map (Eq.~2 of \Kol{}) from the \eSoftB{}. 
		
		The calculation of the mask $\map{M}$ has been also changed. To ensure that low exposed pixels in the CCD gaps and at the edges of the \acisi{} are removed in a consistent manner for all energy bands, we reduce the dimensions of the chip region of each of the four \acisi{} chips%
			\footnote{
				The original dimensions of all chip regions are taken from the FOV-Region-File \url{acisf0xxxx_repro_fov1.fits}, where \url{xxxx} represents the observation ID.
			}
		by $8\,\%$. 
		In this way the FOV mask takes only the inner $\approx85\,\%$ area of each chip into account, which reduces the total  FOV area%
			\footnote{
				Note that \acisi{}'s chip areas are overlapping due to the dithering motion of \chandra{}. 
				Hence a $15\,\%$ reduction of the chip area leads only to a $\approx12\,\%$ reduction of the FOV area.
			}
		by $\approx12\,\%$. 
		In addition, we exclude all pixels with the exposure time less than $2.5$~ks.
		The resulting mask is close to   the FOV mask from \Kol{}, which was created by using a fractional threshold of $63\,\%$ of the peak value of the exposure map in the \eSoftB{}.

		\subsubsection{Count maps} \label{s:C_maps}
		We are now using the \emph{instrumental-background-subtracted count map} $\map{C}$ (counts) of each observation:
			\begin{align} \label{eq:NetCtsMap}
				\map{C} = \map{M} \cdot \left( \map{C}^\mathrm{Total} - \map{C}^\mathrm{BKG} \right) \text{ .}	
			\end{align}
		Here, $\map{C}^\mathrm{Total}$ is the \emph{total-count map}  and $\map{C}^\mathrm{BKG}$ the \emph{instrumental-background map}, which is {computed from}   \chandra{}'s \acisi{} stowed background map%
				\footnote{\label{StowedBKG}\url{http://cxc.harvard.edu/contrib/maxim/acisbg/}}
		($\map{C}^\mathrm{Stow}$) as follows:
			\begin{align} \label{eq:BKG}
				\map{C}^\mathrm{BKG} = \map{C}^\mathrm{Stow} \cdot S
				\text{ ,}	
			\end{align}
		where the rescaling factor $S$ of $\map{C}^\mathrm{Stow}$ is defined as:
		\begin{align} \label{eq:BKG_res}
				S =   \dfrac{\Sigma_{i,j} \; M \cdot C_{\ePart}^\mathrm{Total} }{\Sigma_{i,j} \; M \cdot C_{\ePart}^\mathrm{Stow} } 
				\text{ .}	
			\end{align}
		For a justification of this method see \citet{Hickox2006} and \Kol{} (section~2.4).
		In \Kol{} we used the total-count map instead because for the stacked fluctuation signal the instrumental-background subtraction was not necessary (see Fig.~D3 of \Kol{}). 
		For computing the \PoSp{} of the mosaic instrumental-background-subtraction becomes critical because of the variations of the instrumental background from observation to observation {(\Aref{a:Bkg})}.
		
		\subsubsection{Removing resolved sources} \label{s:ResoPoSo}
		We simplify the {procedure used to remove} \rPS{} in comparison to the 
		\Kol{} (section~2.2.1).
		Now, all point sources are removed with the same  circular exclusion region of the radius of $20\arcsec$.
		The value of the exclusion  radius was chosen based on the results of \Kol{} and is about half the size of the average radius used before ($\approx44\arcsec$).
		This  method increases the  area {remaining after the \rPS{}  removal} by $\sim18\,\%$  in comparison to the method of  \Kol{}, which leads to a slight increase of the S/N of our fluctuation measurement. % (\Fref{PS_DefMsk}).
		Most importantly, it minimizes the selection bias, which arises from the fact that a circular exclusion area of a point source could potentially mask out photons from other  CXB components and therefore alter their correlation signal.
		We compare and discuss the methods of this work and of \Kol{} in \Aref{a:Data}.
		
		Extended sources are removed with the same procedure as in \Kol{}.
		Also see \Aref{a:LSS_rES_frac} for a quantitative justification.
		
		{As in \Kol{} we define two masks:
		\begin{enumerate}
		\item \emph{Default mask}, in which all resolved sources (point sources and extended sources) are masked out on the image (\Fref{MSK_RS10});
		\item \emph{Special mask}, in which only point sources are masked out, while extended sources are retained on the image (\Fref{MSK_RS09})
		\end{enumerate}}

		\subsubsection{Luminosity function of \GCG{}} \label{s:XLF}
		%-----------------------------------------------------------
		We are now using the  most recent XLF of \citet{Pacaud2016}.
		It is based on $100$ bright \GCG{} detected in the XXL survey with fluxes above $\sim4\times10^{-14}\,\mathrm{erg\,cm^{-2}\,s^{-1}}$. 
		The sample covers redshifts up to $z\sim1$ and luminosities down to $\sim1\times10^{42}\,\mathrm{erg\;s^{-1}}$ and 
		the resulting XLF does not shown any redshift evolution.
		We approximate the XLF with a Schechter function: $\Phi(L) = A \, (L/L_0)^{-\alpha} \, \exp(-L/L^\ast)$
		with $A = 8.94\times10^{37}\,h^5\,\mathrm{Mpc^{-3}\,erg^{-1}\,s}$,
		$\alpha = 2.01$,  $L_0 = 10^{43}\,h^{-2}\,\mathrm{erg\,s^{-1}}$, 
		and $L^\ast = 1.72\times10^{44}\,h^{-2}\,\mathrm{erg\,s^{-1}}$.
		We rescale the XLF by a factor of $0.51$ to match the predicted \LogNLogS{} of the XLF with the observed \LogNLogS{} of extended sources in XBOOTES (\Ken{}, Table~1) at $S_{\eSoft}=3\times10^{-14}\,\mathrm{erg\,cm^{-2}\,s^{-1}}$,
		an approximate XBOOTES flux limit for extended sources.
		To exclude very low-mass DMHs, we impose a lower ICM temperature limit of $T=0.4\,\mathrm{keV}$ ($\approx5\times10^6\,\mathrm{K}$).
		At the median redshift of the unresolved population this corresponds to a lower limit in luminosity of $L_{\eSoft}\approx 2 \times10^{41}\,\mathrm{erg\,s^{-1}}$ and in DMH mass of $M_{500}\approx 6 \times 10^{12}\,\mathrm{M_{\sun}\,}h^{-1}$ (\Tref{tab:APEC_fit}),
		using the luminosity-temperature scaling relation of \citet{Giles2015} and the the mass-temperature scaling relation of \citet{Lieu2016}.
		It also leads to a flattening of the cumulative \LogNLogS{} at fluxes below $\sim10^{-17}\,\mathrm{erg\,cm^{-2}\,s^{-1}}$.
		Apart from this, we do not change the procedure used in \Kol{} to compute the surface brightness, and redshift and luminosity distributions (Eqs.~9-11, \Kol{}) of the unresolved population of \GCG{} in XBOOTES.
		
		With the rescaled XLF of \citet{Pacaud2016} we estimate a surface brightness of $\approx3.1\times10^{-13}\,\mathrm{erg\,cm^{-2}\,s^{-1}\,deg^{-2}}$ for the unresolved \GCG{},
		which corresponds to $\approx7\,\%$ of the total unresolved emission of XBOOTES (using Table~3 of \Kol{}).

	\subsection{Mosaic} \label{s:MSC}
	%-----------------------------------------------------------
		\begin{figure}
		\begin{center} % .
			% set f="MRG_Pix32_Band0.5_2.0_MSK03_FOV092_DF03_RS10_neg"
			% convert ${f}.jpg eps2:${f}.ps
			\resizebox{\hsize}{!}{\includegraphics{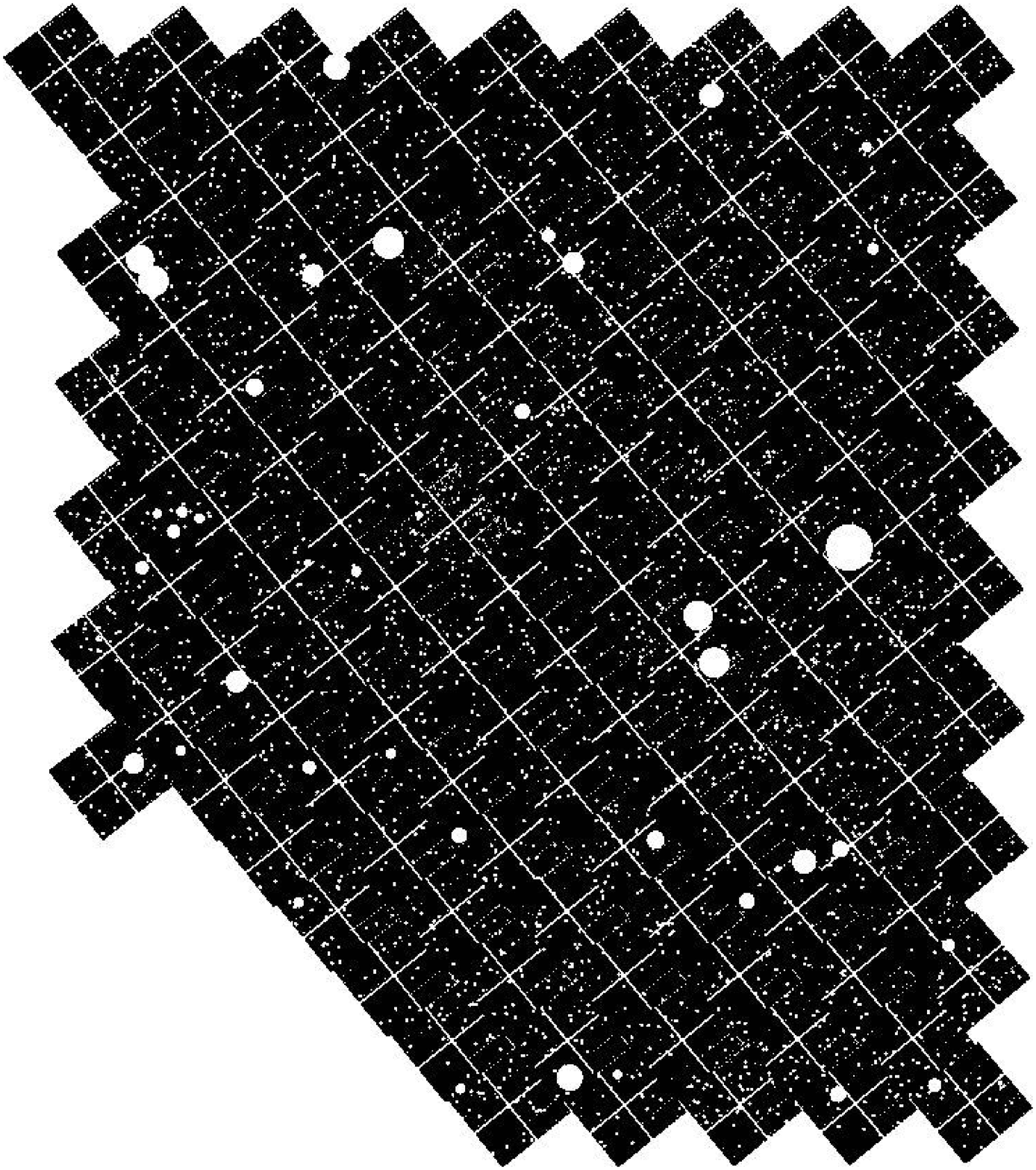}} 
			\caption{\label{f:MSK_RS10} 
				Default mosaic mask, which is used to remove all resolved point and extended sources from the flux mosaic (\Eref{eq:F_x}). 
				}
		\end{center}		
		\end{figure}
		
		\begin{figure}
		\begin{center} % .
			\resizebox{\hsize}{!}{\includegraphics{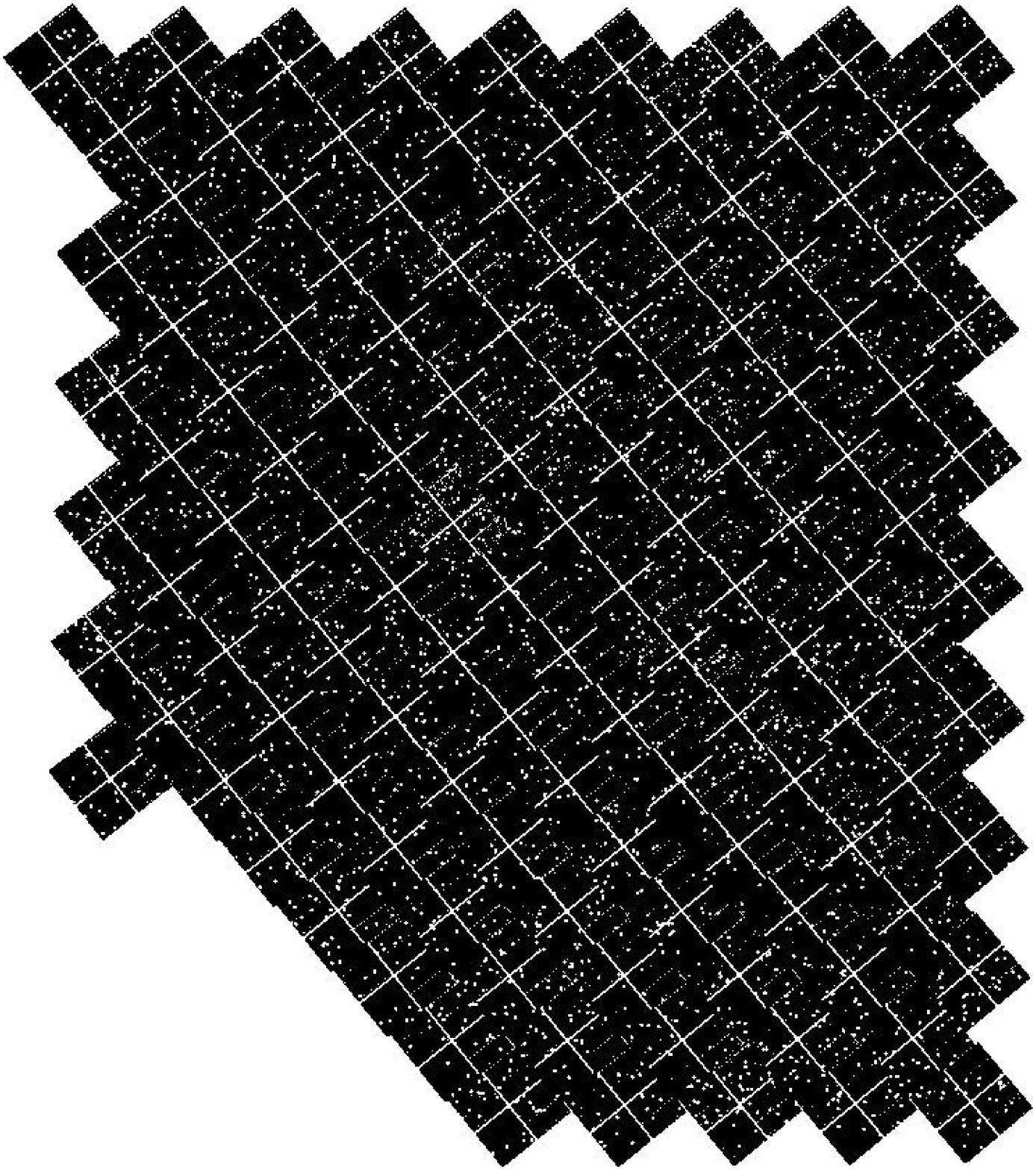}} 
			\caption{\label{f:MSK_RS09} 
				Special mosaic mask, which is used to remove only \rPS{} from but retain \rES{} in the flux mosaic (\Eref{eq:F_x}).
				}
		\end{center}		
		\end{figure}
		
		\subsubsection{Construction} \label{MSC:const}
		In order to measure CXB fluctuations at angular scales larger than \acisi{}'s FOV ($>17\arcmin$), we have to analysis the mosaic image of 126 \acisi{} observations.
		At first we construct the mosaics $\msc{E}$, $\msc{M}$, and $\msc{C}$ out of the individual exposure maps ($\map{E}$), masks ($\map{M}$), and background-subtracted count maps ($\map{C}$), respectively, for each energy band. 
		Note that before the mosaics are constructed the exposure and count map of each observation are multiplied with its FOV mask (\Sref{s:ETM_MSK}).
		With these mosaics we compute the \emph{flux mosaic} for a given energy band as:
			\begin{align} \label{eq:F_x}
				\msc{F} = \msc{M} \cdot \dfrac{ \msc{C} }{ \msc{E} } \text{ .}	
			\end{align}
		Hence, the fluctuation mosaic is computed as:
			\begin{align} \label{eq:FlucMap}
				\delta\msc{F} =  \msc{F} - \left\langle \msc{F} \right\rangle \text{ ,}
			\end{align}			
		where the \emph{mean flux mosaic} is:
			\begin{align} \label{eq:MeanFlx}
				\langle \msc{F} \rangle = \msc{M} \cdot \dfrac{\Sigma_{i,j} \, F }{\Sigma_{i,j} \, M } \text{ .}
			\end{align}
		Here and in the following, blackboard bold characters (\msc{M}) represent mosaics, while bold italic characters (\map{M}) represent individual maps, and italic characters ($M$) represent individual image pixels of either a mosaic or a map (depending on the context).
		Our construction method ensures that overlapping regions of adjacent observations ($\sim5\,\%$) are properly taken into account (also see \Aref{a:OL}).
		
		Note that the fluctuation mosaic is created with an image-pixel-binning factor of $b=32$, while individual \emph{fluctuation maps} ($\delta\map{F}$, Eq.~8 of \Kol{}), which are used to compute the {\sPoSp{}}, are created at the highest possible angular resolution of \acisi{} (image-pixel-binning of $b=1$).
		Hence, the former has a image-pixel-size of $15.744\arcsec$, while the latter have an image-pixel-size equal to the chip-pixel-size
		of \acisi{}\footnote{\label{ACISI_Pix}\url{http://cxc.harvard.edu/proposer/POG/html/chap6.html\#tab:acis_char}},
		which is $\Delta p=0.492\arcsec$.
		Using a reduced angular resolution for the mosaic image makes our data analysis significantly faster and more manageable. 				
		To optimize the discrete Fourier transform computations of our analysis, 
		the fluctuation mosaic is embedded in a squared image of $837\times837$ image pixels ($\approx3.66\times3.66\,\mathrm{deg^2}$).
		Fluctuation maps are  embedded in a squared image of $2\,900\times2\,900$ image pixels ($\approx23.8\times23.8\,\mathrm{arcmin^2}$), which is large enough to contain an entire \acisi{} FOV.
		
		\subsubsection{Solid angle and flux}   \label{MSC:stat} \label{s:MSC_Area} \label{s:MSC_flux}
		The {solid angle}  of the mosaic mask ($\msc{M}$) is computed as 
			\begin{align} \label{eq:Omega}
				\Omega = (b \cdot \Delta p)^2 \cdot \left(\Sigma_{i,j} \, M \right) \text{ ,}
			\end{align}
		The total {solid angle covered by  the mosaic image} is $\approx8.7\,\mathrm{deg^2}$, of which about $5\,\%$  are {covered by} two or more observations.
		When we apply our \emph{default mask} shown in \Fref{MSK_RS10}, which removes all resolved sources, the remaining area reduces by about $5\,\%$ down to $\approx8.3\,\mathrm{deg^2}$.
		For the \emph{special mask} shown in \Fref{MSK_RS09}, which {is produced from the default mask by retaining all \rES{},} the remaining area is $\approx8.4\,\mathrm{deg^2}$.
		The average exposure time is $\approx4.5$~ks (overlap corrected),
		which is $\sim5\,\%$ higher than in \Kol{} due to the changes in the data processing (\Sref{s:ETM_MSK}). 
		
		The average surface brightness  of the flux mosaic ($\msc{F}$) is 
		$0.77\pm0.01\,\mathrm{counts\;s^{-1}\,deg^{-2}}$ in the \eSoftB{} 
		after removing all resolved sources (default mask).
		This corresponds to 
		$7.5\pm0.1\,\times10^{-12}\,\mathrm{erg\,cm^{-2}\,s^{-1}\,deg^{-2}}$  % p,reform(SrfB_eBand_cgs[*,0,0])
		in physical units, which is $\sim5\,\%$ smaller than in \Kol{} due to the changes in the data processing (\Sref{s:ETM_MSK}). {This discrepancy characterizes the amplitude of the systematic uncertainty of the absolute flux measurements in this work and in    \Kol{}.}
		If all \rES{} are retained on the image (special mask), the average surface brightness increases  by $\sim4\,\%$ to 
		$0.80\pm0.01\,\mathrm{counts\;s^{-1}\,deg^{-2}}$, % p,reform(SrfB_eBand[*,1,0,1])
		which corresponds to 
		$7.8\pm0.1\,\times10^{-12}\,\mathrm{erg\,cm^{-2}\,s^{-1}\,deg^{-2}}$. % p,reform(SrfB_eBand_cgs[*,0,1])
		The difference between the flux computed with the special and default mask gives us the combined surface brightness of all \rES{},
		which is $2.8\pm0.8\,\times10^{-2}\,\mathrm{counts\;s^{-1}\,deg^{-2}}$. % p,reform(SrfB_eBand_rES[*,0])
		This corresponds to $2.0\pm0.6\,\times10^{-13}\,\mathrm{erg\,cm^{-2}\,s^{-1}\,deg^{-2}}$  % p,reform(SrfB_eBand_rES_cgs[*,0])
		using the best-fit spectral model from \Sref{s:LSS_Espec}.

		\begin{figure}
		\begin{center} % .r p67_MSC_h_TalkPlots.pro
			\resizebox{\hsize}{!}{\includegraphics{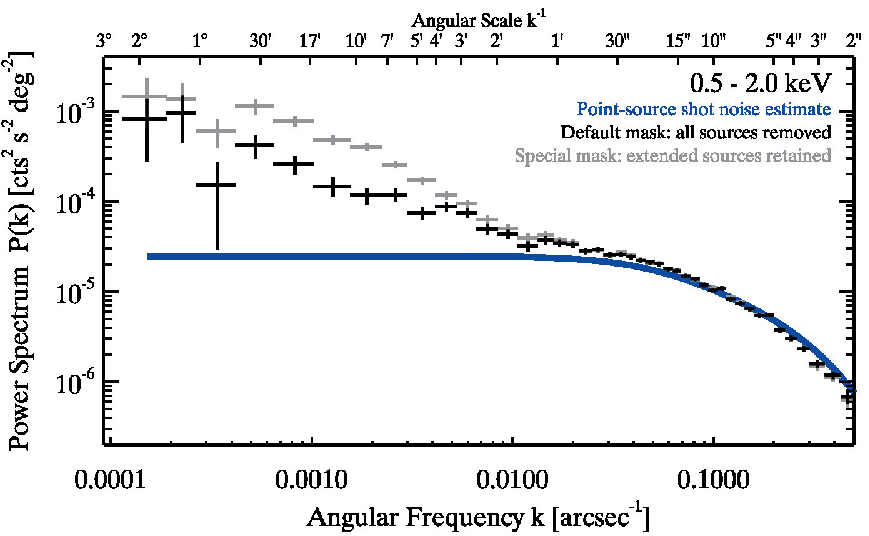}}
			\caption{\label{f:PS_eSoft} 
				The measured \PoSp{} of the surface brightness fluctuations of the CXB in the \eSoftB{}
				for the default mask (black, \Fref{MSK_RS10}), where all resolved sources are removed,
				and for the special mask (gray, \Fref{MSK_RS09}), where in comparison to the default mask all \rES{} are retained.
				The blue curve represents our observational estimate of the \PoSoSN{} (\Sref{s:PoSoSN}) multiplied by the PSF-smearing model (\Aref{a:PSFsmear}). 
				The \PoSpa{} are adaptively binned. 
				}
		\end{center}
		\end{figure}
	%----------------------------------------------------------------------------------------------------------------------
	\section{Power spectrum of CXB fluctuations} \label{s:Fluc}
	%---------------------------------------------------------------------------------------------------------------------- '
				
	\label{s:Def} \label{s:combinedPS}
	%-----------------------------------------------------------
		We {use the same  formalism as} in \Kol{} (section~4.1) to compute the \PoSpa{} of {the mosaic image} ($\delta\msc{F}$, \Eref{eq:FlucMap}) {and of the average \PoSp{}} of individual observations ($\delta\map{F}$, Eq.~8 of \Kol{}).
		We will refer to the former  as \emph{\mPoSp{}} ($P_\mathrm{M}(k)$) and to the latter  as  \emph{\sPoSp{}} ($P_S(k)$).
		
		The \mPoSp{} covers  angular scales  from $\sim32\arcsec$  up to  about $3^\circ$ (angular frequencies $\sim 10^{-4} - 0.0313\InvArcSec$), where the maximal angular scale is determined  by the geometry of the XBOOTES survey.
		The \sPoSp{} covers   angular scales from $\sim1\arcsec$ up to  about $17\arcmin$ (angular frequencies $\sim 10^{-3} - 0.300\InvArcSec$), with the maximal angular scale determined by \acisi{}'s FOV. 
		The minimal angular scale in both cases is defined by the Nyquist frequency ($k_\mathrm{Ny} = ( 2\,b\,\Delta p )^{-1}$), which depends on the image-pixel-binning used for the mosaic image ($b=32$) and for individual observations ($b=1$, \Sref{MSC:const}).
		We combine the mosaic and stacked power spectra in order to cover the entire range of angular scales from $\sim1\arcsec$ to $\sim 3^\circ$.
		This is done after  subtraction of the \PhSN{} (\Aref{a:PhoSN}).
		The two spectra are combined at the half of the Nyquist frequency of the \mPoSp{}   
		$k_\mathrm{C} = k_\mathrm{Ny}^\mathrm{Mosaic} / 2 \approx ( 63\arcsec )^{-1} \approx 0.016\InvArcSec$.
		This choice of $k_\mathrm{C}$ is justified in  {\Aref{a:PSs}}.
		The characteristics of all three power spectra are summarized in \Tref{tab:PSs}.
		For simplicity, in the following we will refer to the \PhSN{} subtracted \cPoSp{} as  \emph{\cxbPS{}}.
		
		\label{s:MeasuredPS}	
		%----------------------------------------------------------------------------------------------------------------------
		
		The  \cxbPS{} in the \eSoftB{} is shown in \Fref{PS_eSoft} for the default and special masks.
		For visualization purposes, we use  adaptive binning in this and following plots.
		All fits of the power spectra were done to the  unbinned data.
		As discussed in detail on \Kol{}, the \cxbPS{} has a significant contribution of the shot noise due to unresolved sources.
		It is shown in \Fref{PS_eSoft} by the thick blue solid curve.
		Clustering and internal structure of unresolved sources leads to the deviations of the \cxbPS{} above the shot noise of unresolved sources.
		In the following, we will refer to any excess power above the shot noise of unresolved {point} sources $P_\mathrm{PSSN}$  as the \emph{\lssPS{}} $P_\mathrm{LSS}(k)$. 
			\begin{align} \label{eq:PS_Meas}
				P(k) = \left(  P_\mathrm{LSS}(k) + P_\mathrm{PSSN} \right) \, W_\mathrm{PSF}(k) 
			\end{align}
		Note, that both $P_\mathrm{LSS}(k)$ and $P_\mathrm{PSSN}$ are affected by the smearing effect of \chandra{}'s point spread function ({$W_\mathrm{PSF}(k)$}, \Aref{a:PSFsmear}), which leads to the decline of the power at high spatial frequencies.

	\subsection{Point-source shot noise} \label{s:PoSoSN}	
	%---------------------------------------------------------------------
		
		In order to study the \lssPS{}  the \PoSoSN{}  needs to be subtracted from the \cxbPS{} (\Eref{eq:PS_Meas}).
		It is an additive, scale-independent component, which arises from the fluctuation of the number of unresolved point-like sources (AGN and normal galaxies) per beam, similar to the \PhSN{} (see section~4.3 of \Kol{} for a more detailed discussion).
		Unlike the \PhSN{}  it is however affected by the PSF-smearing.
		
		In theory, the amplitude of the shot noise of unresolved {point} sources can be straightforwardly computed form the \LogNLogS{} distribution of the unresolved sources (e.g. Eq.~18 of \Kol{}).
		However in practice, the theoretical prediction is subject to a number of uncertainties (section~5.1 of \Kol{}), of which one of the most significant is  conversion from physical to instrumental units.
		For this reason we estimate the amplitude of the \PoSoSN{} directly from the \cxbPS{} itself.
		To this end, we compute the average power within the $5\arcsec - 10\arcsec$ angular scale range
		and correct it for the PSF-smearing ($W_\mathrm{PSF}$, \Aref{a:PSFsmear}):
		\begin{align} \label{eq:PS_Ave}
			\langle P_{(k_1,k_2)} \rangle = \frac{ \Sigma_{k_1}^{k_2} P(k)}{\Sigma_{k_1}^{k_2} W_\mathrm{PSF}(k)}
			\text{ .}	
		\end{align}
		where, $k_1$ and $k_2$ are the lower and upper limit of the considered angular scale range.
		From this calculation we obtain the \PoSoSN{} level of  $2.5\pm0.1\times 10^{-5}\,\mathrm{(cts\,s^{-1})^2\, deg^{-2}}$ in the \eSoftB{}. 
		This value is consistent with the one obtained in \Kol{}.
		The so computed \PoSoSN{} contribution is shown  in \Fref{PS_eSoft} by the blue curve. % thick {horizontal} line.
		
		In \Fref{PoSoSN_eSoft} we compare this value (gray horizontal bar) with theoretical predictions based on  different \LogNLogS{} of extragalactic point sources in the \eSoftB{} from the literature.  
		In this figure, the theoretical shot noise level is plotted as function of the photon index of the power law, assumed in  converting the energy flux to instrumental units.
		We can see that theoretical predictions agree with our measurement if we assume the effective average photon index of unresolved extragalactic point sources of $\Gamma\sim1.6-1.7$.
		These values are somewhat lower than the best fit value of $\Gamma=1.73\pm0.03$ obtained in \Kol{} (section~3.1.3) from the power law fit to the energy spectrum of the unresolved extragalactic emission in XBOOTES in the $0.5-10.0\,\mathrm{keV}$ band.
		We consider this agreement satisfactory, given the number of uncertainties involved in the measurement of the \PoSoSN{} level and in the theoretical calculation.
		Finally we note that Galactic absorption ($\NH=10^{20}\,\mathrm{cm^{-2}}$) is always taken into account for the power law model.

		\begin{figure}
		\begin{center} % .r p67_MSC_h_TalkPlots.pro
			\resizebox{\hsize}{!}{\includegraphics{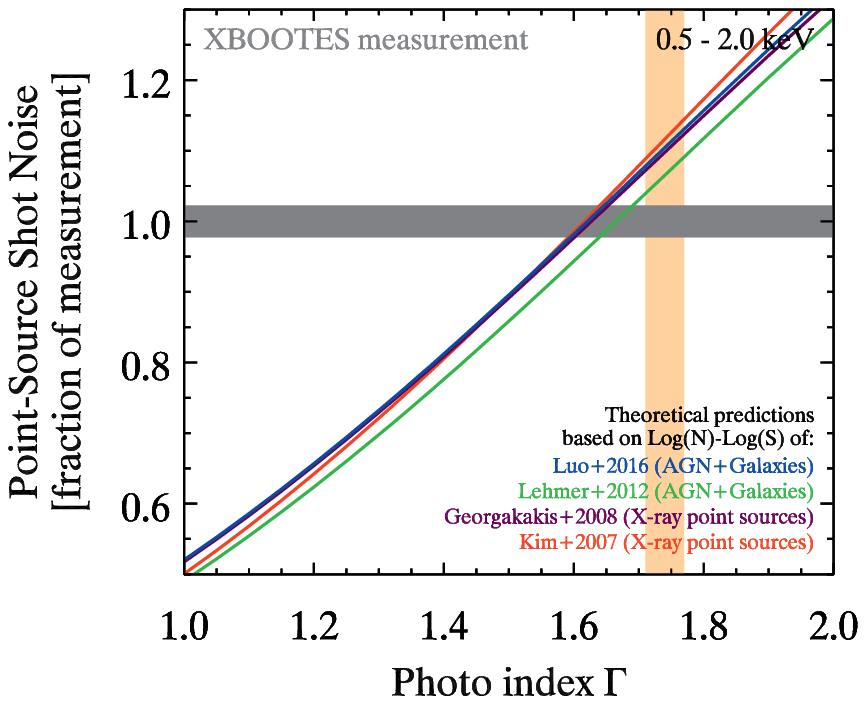}}  
			\caption{\label{f:PoSoSN_eSoft} 
				Comparison of the measured \PoSoSN{} level in the \cxbPS{} in the \eSoftB{} (gray horizontal bar, which thickness corresponds to  one standard deviation)  with theoretical predictions based on  different \LogNLogS{} of extragalactic point sources from the literature 
				\citep[solid curves, computed according to Eq.~18 of \Kol{},][]{Kim2007,Georgakakis2008,Lehmer2012,Luo2017}
				as a function of the photon index of the power law, {assumed in converting the energy flux to} instrumental units. 
				The vertical shaded region shows the $1\sigma$ confidence interval for the slope of the power law -- best fit  in the $0.5-10.0\,\mathrm{keV}$ band to the unresolved extragalactic emission in XBOOTES.
				See {\Sref{s:PoSoSN}} for discussion.
				}
		\end{center}
		\end{figure}

	%---------------------------------------------------------------------
		\begin{figure*}
		\begin{center} % .r p67_MSC_h_TalkPlots.pro
			\includegraphics[width=0.49\textwidth]{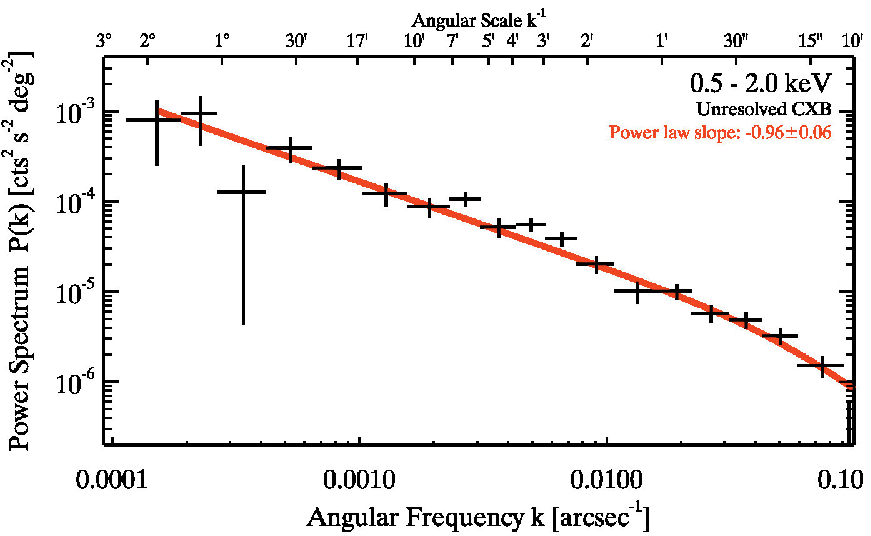}
			\hfill
			\includegraphics[width=0.49\textwidth]{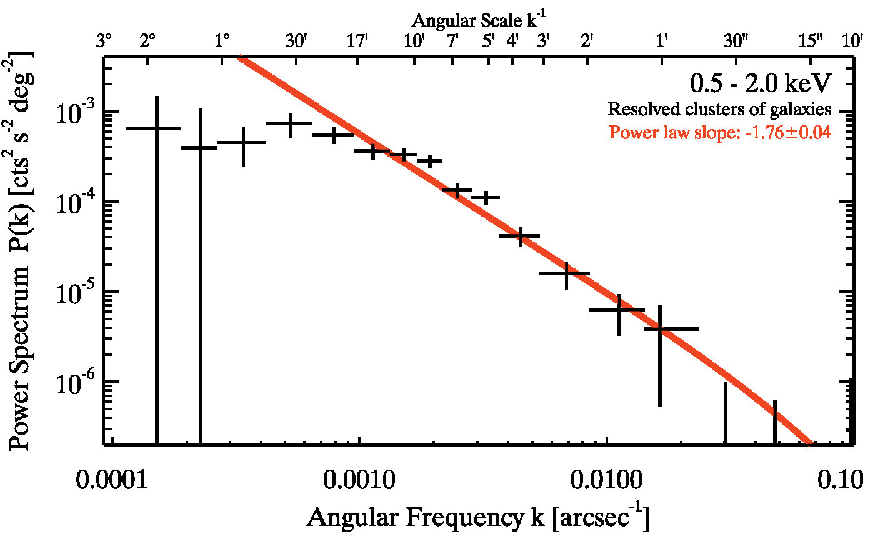}
			\caption{\label{f:PS_eSoft_NoPSSN}  \label{f:PS_eSoft_rES} 
				The {\PoSp{}} of the unresolved CXB (left panel) and of \rES{} (right panel) in the \eSoftB{}.
				The red lines show the best-fit power law model in the angular scale range $10\arcsec - 20\arcmin$ multiplied by the PSF-smearing model (\Aref{a:PSFsmear}).
				The \PoSpa{} are adaptively binned. 
				}
		\end{center} 
		\end{figure*}		
	%---------------------------------------------------------------------
	\section{The LSS power spectrum} \label{s:LSS_PL}
	%---------------------------------------------------------------------
		We defined the \lssPS{} ($P_\mathrm{LSS}(k)$) as the excess power above the shot noise of unresolved point sources ($P_\mathrm{PSSN}$).  
		$P_\mathrm{LSS}(k)$ is computed by subtracting {$P_\mathrm{PSSN}$} from the \cxbPS{} and describes structure and correlation properties of unresolved sources.
		We can also compute the \PoSp{} of resolved \GCG{}%
			\footnote{\label{footn:confirmed_gc}
				We refer to resolved extended sources as to \GCG{} bearing in mind that  only $\approx72\,\%$ (31) of  \rES{} detected in XBOOTES field are confirmed \GCG{} \citep{Vajgel2014}.
			}
		$P_\mathrm{rCG}(k)$ by subtracting the \cxbPS{} computed with the default mask (all resolved sources masked out) from that of the special mask (only resolved point sources masked out).
		It  describes the internal structure and cross correlation of resolved \GCG{}. 
		
		In \Fref{PS_eSoft_NoPSSN} we present our measurement of the \lssPS{} of the unresolved CXB  and of resolved \GCG{}.
		In both cases the power spectra have a power law shape  in a rather broad range of angular frequencies exceeding two orders of magnitude. However, the slopes of the power spectra are significantly different, with the unresolved \cxbPS{} being significantly flatter.
		The power law fits $P(k)\propto k^{-\alpha}$  to these spectra gives best-fit values of slope $\alpha=0.96\pm0.06$ and $\alpha=1.76\pm0.04$ for the unresolved CXB and resolved \GCG{}, respectively. 
		Note, that the best-fit values depend on the survey area and depth.
		There  were determined via an $\chi^2$ minimization \citep[using MPFIT from][]{MPFIT} in the $10\arcsec - 20\arcmin$ angular scale range.
		Also, the \PoSp{} of resolved \GCG{} has a clear flattening at low frequencies corresponding to angular scales larger than  $\sim20\arcmin-30\arcmin$.
		
	%---------------------------------------------------------------------
	\subsection{Resolved \GCG{}} \label{s:PS_rES}
	%---------------------------------------------------------------------
		
		%---------------------------------------------------------------------
			\begin{figure*}
			\begin{center} % .r p67_MSC_h_TalkPlots.pro
				\includegraphics[width=0.43\textwidth]{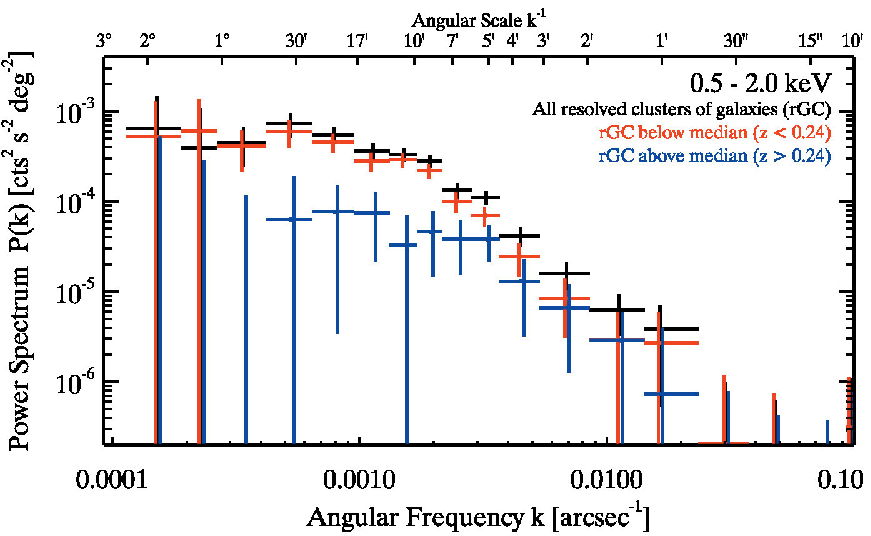}
				\hfill
				\includegraphics[width=0.43\textwidth]{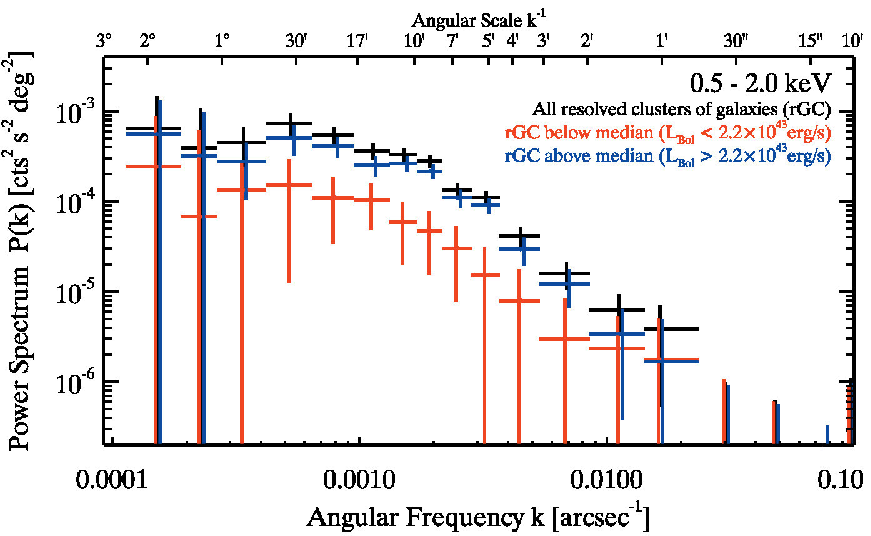}
				\caption{\label{f:PS_eSoft_NoPSSN_Zgroups}\label{f:PS_eSoft_NoPSSN_Lgroups}
					The \PoSp{} of resolved \GCG{} for different redshift  (left) and luminosity groups (right). 
					}
			\end{center}
			\end{figure*}		
		
			\begin{figure*}
			\begin{center} % .r p67_MSC_h_TalkPlots.pro
				\includegraphics[width=0.40\textwidth]{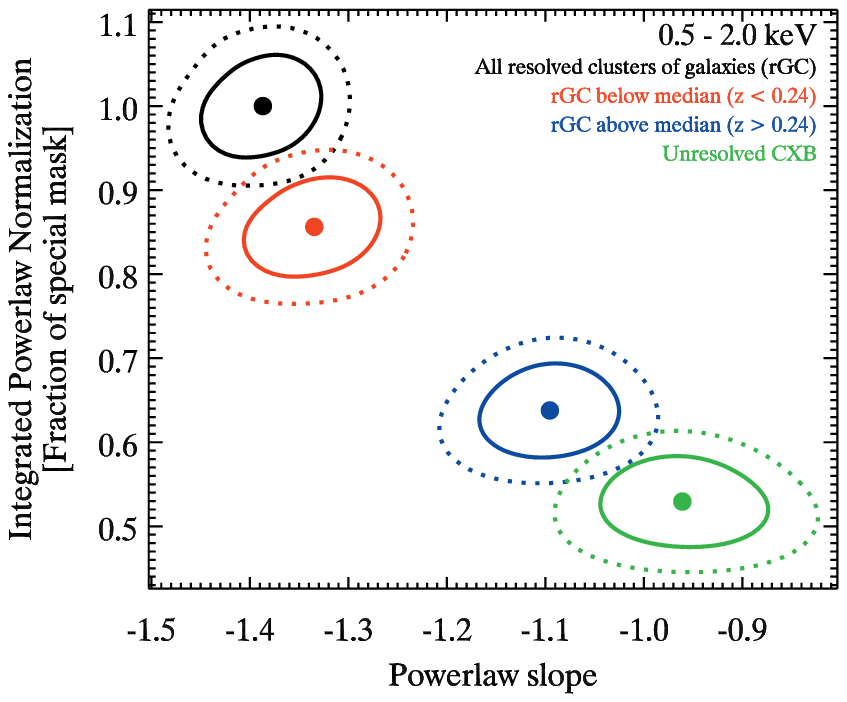}
				\hfill
				\includegraphics[width=0.40\textwidth]{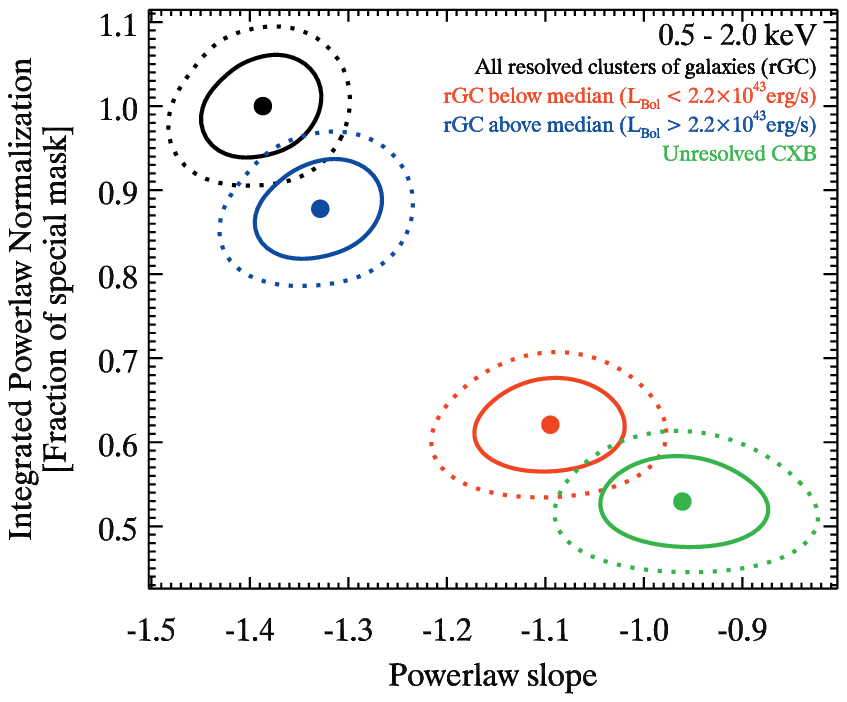}
				\caption{\label{f:Zgroups_Fit} \label{f:Lgroups_Fit} 
					Best-fit parameters for the power law fit to the \PoSp{} of different redshift (left) and luminosity groups (right) of resolved \GCG{}.
					The corresponding power spectra are shown in \Fref{PS_eSoft_NoPSSN_Zgroups}.
					Solid ellipses represent one standard deviation and dashed ellipses show the $90\,\%$ confidence level of a two parameter fit.
					The power law normalization is obtained by integrating the model over the angular scale range $10\arcsec - 20\arcmin$ and is shown in the units of $\mathrm{counts^2\,s^{-2}\,deg^{-3}}$.
					}
			\end{center}
			\end{figure*}		
		%---------------------------------------------------------------------
		\subsubsection{Redshift and luminosity dependence} \label{s:LSS_redshift} \label{s:LSS_lum}		
		%---------------------------------------------------------------------
		Thanks to the work of \citet{Vajgel2014}, we know the redshifts and  luminosities  of $\approx72\,\%$ (31) of  \rES{} detected in the XBOOTES field.
		Their median values are $z\approx0.24$ and $L_{\eBol{}} \approx 2.2\times10^{43}\,\mathrm{erg\;s^{-1}}$, respectively.
		{With these values, we estimate a DMH mass of $M_{500}\approx1.0\times10^{14}\,\mathrm{M_{\sun}}\,h^{-1}$ and an ICM temperature of $T\approx2.2$~keV, using the luminosity-mass relation of \citet{Anderson2015} and the luminosity-temperature relation of \citet[see \Sref{s:Espec_Theo} for the spectral model]{Giles2015}, respectively.}
		Based on this we compute the \eSoftB{} luminosity of $L_{\eSoft{}} \approx  1.2\times10^{43}\,\mathrm{erg\;s^{-1}}$.
		
		We use the median values to divide the sample of resolved \GCG{} into two groups in redshift and luminosity, separated by their median values, and plot their power spectra in \Fref{PS_eSoft_NoPSSN_Zgroups}.
		In \Fref{Zgroups_Fit} we plot best-fit parameters of the {LSS power spectra ($P_\mathrm{LSS}(k)$)} approximation with the power law model.
		In this calculation we excluded resolved extended sources without redshift and luminosity information. 
		These figures demonstrate that  the \PoSp{} of resolved \GCG{} is dominated by the nearby or the most luminous objects.		
		
			\begin{figure}
			\begin{center} % .r p67_MSC_h_TalkPlots.pro
				\resizebox{\hsize}{!}{\includegraphics{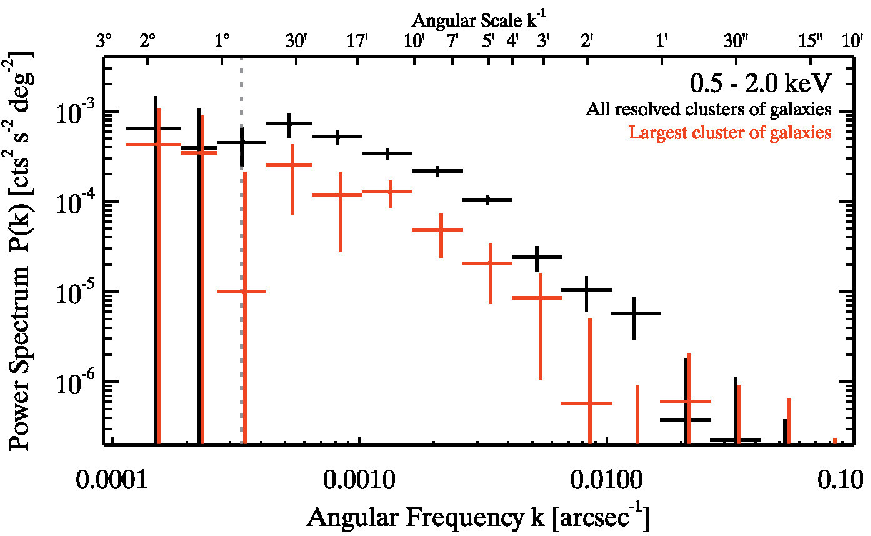}} 
				\caption{\label{f:LSS_Large} 
					{The \PoSp{} in the \eSoftB{} for all resolved \GCG{} (black, same as right panel of \Fref{PS_eSoft_NoPSSN}) and for only the largest cluster (red).}
					The gray dotted vertical line indicates the diameter of the circular exclusion area of the largest cluster.
					}
			\end{center}
			\end{figure}
			
		%---------------------------------------------------------------------
		\subsubsection{Low frequency break} \label{s:LSS_Break}	
		%---------------------------------------------------------------------
		There is a definitive low frequency break in the \PoSp{} of resolved \GCG{}  (right panel of \Fref{PS_eSoft_rES}) at  angular scales of $\sim20\arcmin-30\arcmin$. 
		At lower frequencies the \PoSp{} becomes flat implying that  there is no correlation between surface brightness variations at the locations separated by  angles larger than $\sim20\arcmin-30\arcmin$.
		In \Aref{a:RND} we demonstrate with randomized observations that the {low frequency break} is real and is not of instrumental origin.
		
		The break location and the shape of the \PoSp{} near the break characterizes the structure of ICM in the largest (in terms of the angular size) cluster of galaxies.
		In the XBOOTES field, this is  XBS06 (Table~1 of \citealt{Vajgel2014}; J142657.9+341201 in Table~1 of \Ken{}) located in the the observation 4224.
		In  \Fref{LSS_Large} we show the \PoSp{} of this cluster of galaxies (red). 
		It was computed as a difference of the power spectra of the entire XBOOTES field with this cluster retained or masked out. 
		In this calculation we set the {radius of the circular exclusion area} for XBS06 to be  $25\arcmin$. 
		
		The \PoSp{} of XBS06 has the low frequency  break at angular scales of $\sim30\arcmin$, similar to the \PoSp{} of all resolved \GCG{}. 
		The redshift of XBS06 is  $z\approx0.128$ \citep{Vajgel2014}. 
		For this redshift,  the angular scale of $\sim30\arcmin$ corresponds to the linear scale of $\sim3\,\mathrm{Mpc}\,h^{-1}$. Given the  size of the galaxy cluster \citep[$R_{500}=(0.59 - 0.65)\,\mathrm{Mpc}\,h^{-1}$,][Table~3]{Vajgel2014},
		this suggest that our fluctuation measurement is sensitive to the ICM structure  up to the radius of $\sim3\times R_{500}$.
	
		This further justifies  the claim of the \Kol{} that  CXB fluctuation analysis can be efficiently used  to study the average ICM structure in the outskirts of \GCG{}, out to their virial radii.
		Such studies are  difficult and expensive to perform with conventional deep pointed observations of individual clusters, especially for  a large number of objects,  and are typically biased towards relatively nearby sources \citep[$z\lesssim0.1$, e.g.][]{Eckert2012,Eckert2016a}.

		%---------------------------------------------------------------------
			\begin{figure*}
			\begin{center} % .r p67_MSC_h_TalkPlots.pro
				\includegraphics[width=0.49\textwidth]{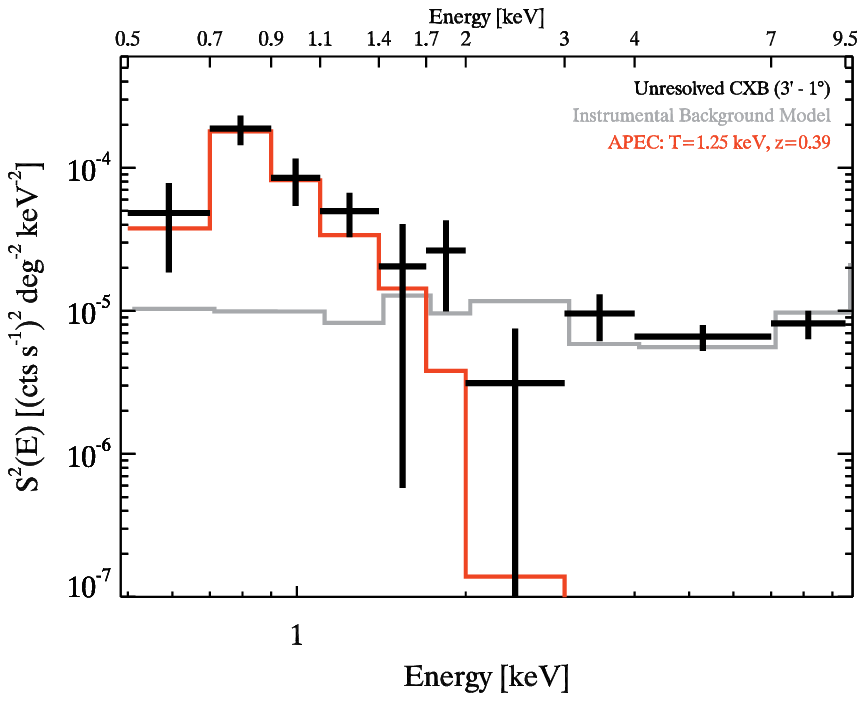}
				\hfill
				\includegraphics[width=0.49\textwidth]{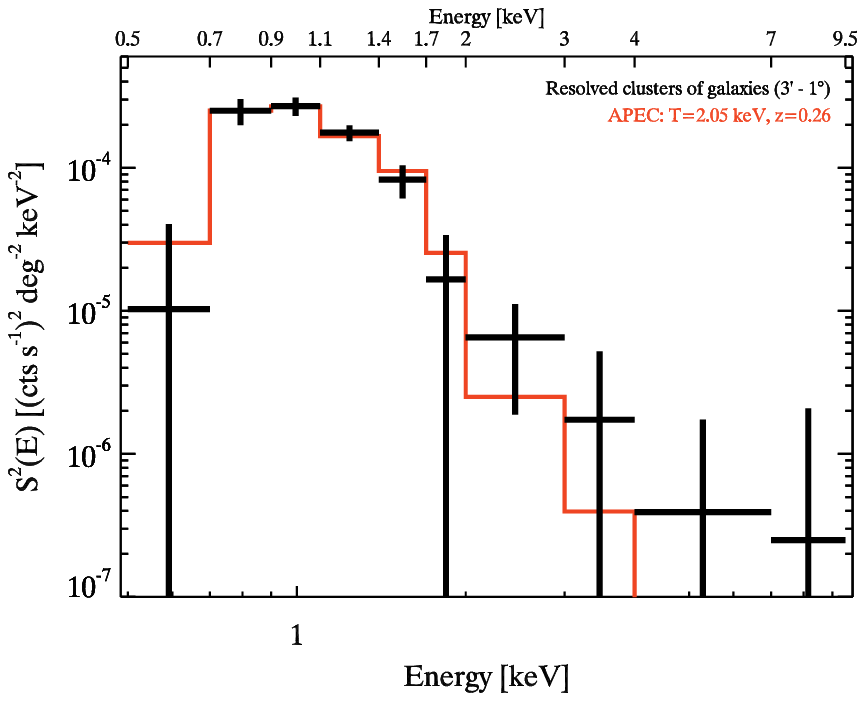}
				\caption{\label{f:LSS_Espec_APEC} 
					The energy spectrum $S^2(E)$ of fluctuations  of the unresolved CXB (left panel) and of resolved \GCG{} (right panel).  
					The best fit  \APEC{} models are shown by red histograms. 
					The gray histogram on the left panel show the residual instrumental background spectrum with the best fit normalization.
					See \Sref{s:LSS_Espec} for details.
					}
			\end{center}
			\end{figure*}
			
			\begin{figure}
			\begin{center} % .r p67_MSC_h_TalkPlots.pro
				\resizebox{\hsize}{!}{\includegraphics{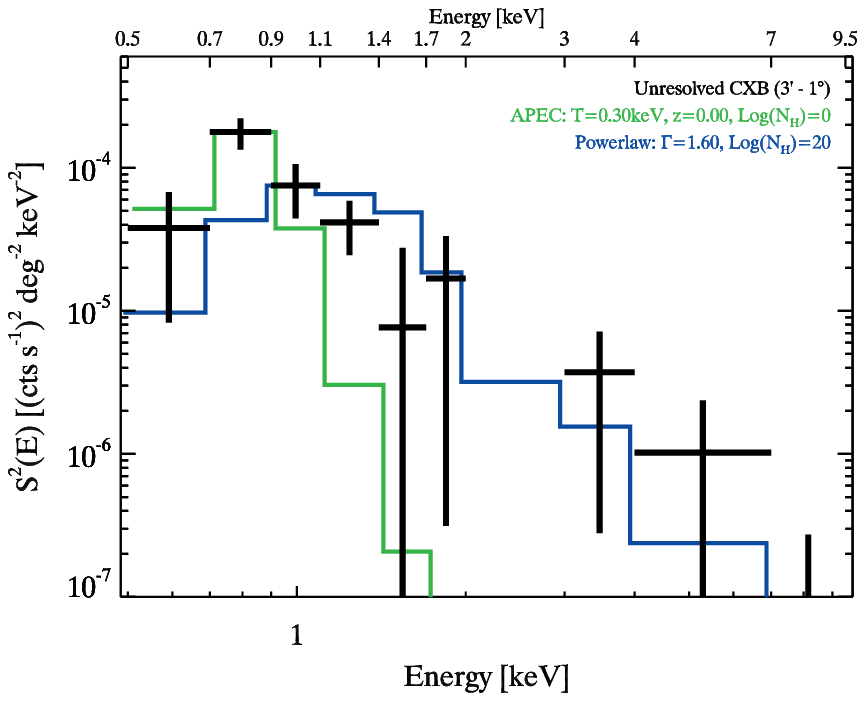}}
				\caption{\label{f:LSS_Espec_others02} 
					The energy spectrum $S^2(E)$ of fluctuations  of the unresolved CXB (after subtracting the instrumental background model)
					in comparison to typical spectral models of AGN and normal galaxies (blue histogram), and of Galactic diffuse emission (green histogram).
					}
			\end{center}
			\end{figure}		
			
		%---------------------------------------------------------------------
		\subsection{Energy spectrum {of fluctuations}} \label{s:LSS_Espec}	
		%---------------------------------------------------------------------
		The energy spectrum of angular fluctuations gives further insights to their origin. We characterize it with the energy dependence of the average power in the  angular scale range of interest. 
		In analogy with  conventional energy spectra, this quantity is further normalized to the square of width of the energy range:
		\begin{align} \label{eq:PS_ESpec}
			S^2(E)= \frac{ \langle P(k_1,k_2) \rangle}{(\Delta E)^2} 
			\text{ ,}	
		\end{align}
		where $\langle P_{(k_1,k_2)} \rangle$ is defined in \Eqref{eq:PS_Ave}. 
		The so defined $S^2(E)$ has the meaning {of} the squared spectral flux {with} units of $\mathrm{(counts\,s^{-1})^2\,deg^{-2}\,keV^{-2}}$.
		For the angular scale range in \Eqref{eq:PS_ESpec} we chose  $3\arcmin - 1^\circ$, which is a compromise between using as wide as possible range to achieve a high S/N and avoiding at the same time angular scales with potential systematic uncertainties%
			\footnote{
			The largest scales are subject to the mask effect (\Aref{a:MCS}) and the smallest scales are compromised by potential uncertainties of the PSF-smearing model {(\Aref{a:PSFsmear})} and of the estimate of the \PoSoSN{} level {(\Sref{s:PoSoSN})}.
			}.
		
		The so obtained \emph{fluctuation spectra} are plotted in \Fref{LSS_Espec_APEC} for the unresolved CXB and resolved \GCG{}.
		Based on theoretical expectations and results of \Kol{} we approximate these spectra with the model of the emission of the optically thin plasma in collisional ionisation equilibrium (\APEC{} model). 
		To preserve the Gaussian statistics of errors we perform fitting in the  squared spectral flux space, i.e. fit the quantity $S^2(E)$. 
		To this end, we construct a grid of models in \xspec{}\footnote{X-Ray spectral fitting package \citep[v12.9.0,][]{XSPEC}.} and then compute $\chi^2$ and find the minimum and confidence intervals outside \xspec{}.
		In the spectral fits we assume Galactic absorption with $\NH=10^{20}\,\mathrm{cm^{-2}}$ and a metallicity of $0.3$ of the solar value.
 
 		There is an obvious hard tail in the  energy spectrum of fluctuations of the unresolved CXB  (left panel of \Fref{LSS_Espec_APEC}).  
 		This is a small residual left because of the imperfect subtraction of the instrumental background (\Sref{s:C_maps} and \Aref{a:Bkg}).
 		To  account for this residual  background contribution we added to the model a component corresponding to the spectrum of the instrumental background which we adopt from \Kol{}  and for which we keep the normalization free during the $\chi^2$ minimization. Note that the instrumental background component is absent in the energy spectrum of fluctuations of resolved \GCG{}, by the method of its construction, as it was computed as a difference of two power spectra having almost exactly same instrumental background contributions. 
		
		The best fit models are shown in \Fref{LSS_Espec_APEC} and their confidence areas are plotted in \Fref{LSS_Espec_APEC_ChiSqr02}.
		The spectra of unresolved CXB fluctuations and of resolved \GCG{} are clearly different, with the former having lower temperature and originating at larger redshift, as it should be intuitively expected. This will be discussed in more detail in the next subsection.
		
		In \Fref{LSS_Espec_others02} we compare the energy spectrum of fluctuations of unresolved CXB with other plausible models.
		The blue histogram shows an extragalactic power law with the photon index of $\Gamma=1.6$, which should be expected from unresolved AGN (\Sref{s:PoSoSN}). 
		Such a spectrum is clearly much harder that the data, which is in good agreement with the \Kol{}. %\At{Something is missing here!}
		Same is true for any power law model with the photon index feasible for AGN and normal galaxies \citep[$0<\Gamma<3$, e.g.][]{Reynolds2014,Ueda2014,Yang2015}.
 			
		% rule out Galactic contribution
		The Galactic diffuse emission can be described on average with an unabsorbed ($\NH=0\,\mathrm{cm^{-2}}$) \APEC{} model with temperatures below $T\sim0.30$~keV \citep[e.g.][]{Lumb2002,Hickox2006,Henley2013}.
		An  \APEC{} model with $T=0.30$~keV is {shown as green histogram} in  \Fref{LSS_Espec_others02}.
		Obviously, it is much softer than the observed spectrum.
		It is possible, however, that the lowest energy {band (\eGal{})} may contain some contribution from the diffuse emission of the Galaxy.
		
		We also tested a more complex, two-component model, where we use an unabsorbed \APEC{} model ($\NH=0\,\mathrm{cm^{-2}}$) to describe the Galactic diffuse emission
		and a power law ($\NH=10^{20}\,\mathrm{cm^{-2}}$) to describe unresolved AGN and normal galaxies.
		It is the same model as used to describe the unresolved CXB emission of XBOOTES (section~3.1 of \Kol{}).
		The best-fit \APEC{} temperature and photon index of such a model are inconsistent with previous measurements for the Galactic and extragalactic components, although errors are large due to poor resolution and low S/N of the energy spectrum.
		Further, our best-fit model of the unresolved CXB emission of XBOOTES (Table~2 of \Kol{}) is excluded by more than $3\sigma$.		
		Due to these reasons, 
		we limit our discussion to single-component models.

		%-----------------------------------------------------------
		% Tables: values
		%-----------------------------------------------------------
			% p67_MSC_h_TalkPlots.pro
			% .r p61_GalClust_LogNLogS_test.pro

			\begin{table}
				\caption{Comparison of the best-fit values of the energy spectrum of CXB fluctuations (\Fref{LSS_Espec_APEC_ChiSqr02}) with theoretical expectations for unresolved and resolved \GCG{}.}
				\label{tab:APEC_fit}
				\begin{center}     % used for centering Table
					\begin{tabular}{l c c}	
						\hline
						\hline
									&   Unresolved$^{(\mathrm{a})}$	& Resolved$^{(\mathrm{b})}$ \\
						\hline
						Redshift z:		& 				& \\
						Best-fit of observation (\Fref{LSS_Espec_APEC})	& $0.39^{+0.10}_{-0.21}$	& $0.26^{+0.28}_{-0.05}$ \\
						Best-fit of simulation$^{(\mathrm{c})}$	& 0.41				& -	\\
						Flux-weighted mean			& 0.35				& 0.23	\\
						Median					& 0.49				& 0.24	\\
						\hline
						Temperature T (keV):	& 				& \\
						Best-fit of observation (\Fref{LSS_Espec_APEC})	& $1.25^{+0.35}_{-0.20}$	& $2.05^{+0.50}_{-0.25}$ \\
						Best-fit of simulation$^{(\mathrm{c})}$	& 1.1				& -	\\
						Flux-weighted mean			& 1.4				& 2.4$^{(\mathrm{d})}$ \\
						Median					& 1.1 				& 2.2$^{(\mathrm{d})}$ \\
						\hline
						\multicolumn{2}{l}{Luminosity $L_{\eSoft{}}$ ($10^{43}\,\mathrm{erg\;s^{-1}}$):} & \\
						Derived$^{(\mathrm{d})}$ from best-fit of observation	
									& $0.3^{+0.3}_{-0.1}$		& $1.0^{+1.1}_{-0.4}$ \\
						% cl=0 & i=0 & p,[l_BestFit_uCXB[0],(lesoft_CL_Arr[0:1,i,1,cl] - l_BestFit_uCXB[0])] * 1d-43; 
						% cl=0 & i=1 & p,[l_BestFit_rCG[0],(lesoft_CL_Arr[0:1,i,1,cl] - l_BestFit_rCG[0])] * 1d-43; 
						Derived$^{(\mathrm{d})}$ from best-fit of simulation	
									& 0.2				& - \\
						Flux-weighted mean	& 0.3				& 1.4 \\
						Median			& 0.2 				& 1.2 \\
						\hline
						\multicolumn{2}{l}{DMH mass $M_{500}$ ($10^{14}\,\mathrm{M_{\sun}}\,h^{-1}$):} & \\
						Derived$^{(\mathrm{e})}$ from best-fit of observation	
									& $0.4^{+0.3}_{-0.1}$		& $1.1^{+0.5}_{-0.3}$ \\
						% cl=0 & i=0 & p,[M_BestFit_uCXB[0],(M500_CL_Arr[0:1,i,0,cl] - M_BestFit_uCXB[0])] * 1d-14;
						% cl=0 & i=1 & p,[M_BestFit_rCG[0],(M500_CL_Arr[0:1,i,1,cl] - M_BestFit_rCG[0])] * 1d-14 ; 
						Derived$^{(\mathrm{e})}$ from best-fit of simulation	
									& 0.4				& - \\
						Flux-weighted mean	& 0.5$^{(\mathrm{e})}$		& 1.0$^{(\mathrm{f})}$ \\
						Median			& 0.3$^{(\mathrm{e})}$ 		& 0.9$^{(\mathrm{f})}$ \\
						\hline
					\end{tabular}
				\end{center}
				(a) Flux-weighted mean and median values are computed using the XLF of {\citet{Pacaud2016} assuming an upper flux limit of $S_{\eSoft}=3\times10^{-14}\,\mathrm{erg\,cm^{-2}\,s^{-1}}$ and a lower temperature limit of $T=0.4$~keV (see \Sref{s:XLF} for details)}.
				(b) Flux-weighted mean and median values are derived from the catalog of \citet[see our \Sref{s:LSS_redshift} for details]{Vajgel2014}.
				{(c) Simulated energy spectrum of unresolved \GCG{} (see \Sref{s:Espec_Theo} for details).}
				(d) Using the luminosity-temperature scaling relation of \citet{Giles2015}.
				(e) Using the mass-temperature scaling relation of \citet{Lieu2016}.
				(f) Using the luminosity-mass scaling relation of \citet{Anderson2015}.
			\end{table}
		%-----------------------------------------------------------			

			\begin{figure}
			\begin{center} % .r p67_MSC_h_TalkPlots.pro
				\resizebox{\hsize}{!}{\includegraphics{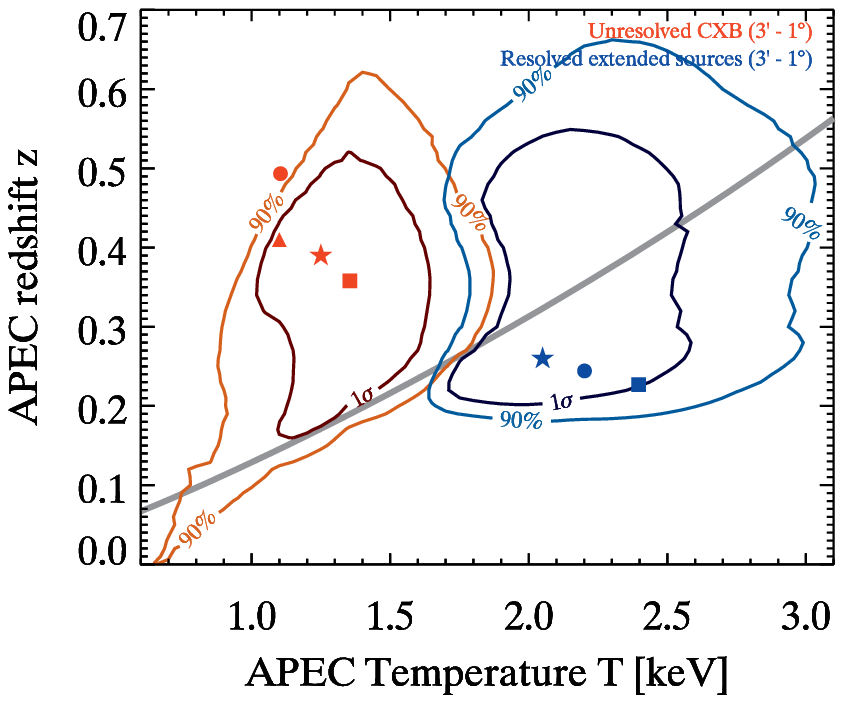}}
				\caption{\label{f:LSS_Espec_APEC_ChiSqr02} 
					Confidence regions for the \APEC{} model parameters for energy spectra of fluctuations of unresolved CXB (red contours) and for resolved \GCG{} (blue contours).
					The stars show corresponding best-fit values,
					the squares and circles show the expected flux-weighted mean and median values for each spectrum, computed as described in \Sref{s:Espec_Theo} and summarized in \Tref{tab:APEC_fit}.
					The red triangle shows the best-fit value for the simulated energy spectrum of unresolved \GCG{}.
					The gray curve shows the maximal redshift at which \GCG{} of a given temperature can be detected in XBOOTES, assuming that their luminosities obey the luminosity-temperature scaling relation of \citet{Giles2015}. 					
					}
			\end{center}
			\end{figure}
			
		%----------------------------------------------------------------------------------------------------------------------
		\subsubsection{Comparison with theoretical expectations}  \label{s:Espec_Theo}
		%----------------------------------------------------------------------------------------------------------------------
		As a consistency check, we  compare the best-fit spectral parameters for resolved \GCG{} with the actually measured values  in the XBOOTES field. 
		We will use the redshift and luminosity measurements from \citet{Vajgel2014} and compute the flux-weighted mean and median values for resolved \GCG{}.  
		Based on these values we compute the expected ICM temperature using the luminosity-temperature scaling relation of \citet[Table~2]{Giles2015}. 
		The result of this calculation is {summarized in \Tref{tab:APEC_fit} and shown in \Fref{LSS_Espec_APEC_ChiSqr02} as a blue rectangle and circle}. 
		As one can see, {they are} quite close to the best-fit value (blue star), although the size of the error region is rather big, due to the limited S/N of the data. 
		
		For unresolved CXB the expected flux-weighted mean and median values
		are derived from the redshift, luminosity and ICM temperature distributions of the unresolved \GCG{},
		which are computed using the XLF $\Phi$ of \citet{Pacaud2016} assuming an upper flux limit of $S_{\eSoft}=3\times10^{-14}\,\mathrm{erg\,cm^{-2}\,s^{-1}}$ and a lower temperature limit of $T=0.4$~keV (see \Sref{s:XLF} for details).
		For the redshift and luminosity distribution we are using Eqs.~(10) and (11) of \Kol{}, while
		for the ICM temperature distribution we are using: 
			\begin{align} 
				\frac{\d S (T)}{\d T} = &
					\int\!\d z \; \Phi\!\left(L_r(T,z),z\right)
						\; L_r^\prime(T,z) 
						\; \frac{\d^2 V(z)}{\d z \d\Omega} \notag \\ & \times
						\; S[L_r(T,z),z] \; \left[1 - f(S[L_r(T,z),z])\right]
					\label{eq:dSdT} \text{,} 
			\end{align} 
		which is the flux production rate per solid angle as a function of temperature.
		$L(T,z)$ and $L^\prime(T,z)$ are the luminosity-temperature scaling relation of \citet{Giles2015} and its first derivative, respectively.
		All other quantities are described in section~3.3 of \Kol{}.
		The result is summarized in \Tref{tab:APEC_fit} and shown in \Fref{LSS_Espec_APEC_ChiSqr02} as the red rectangle and circle.
		
		One can see that the flux-weighted mean (red square in \Fref{LSS_Espec_APEC_ChiSqr02}) is located within $1\sigma$ contour while the median value (red circle) is outside the $90\,\%$ confidence area (but still within $2\sigma$ contour). 
		This discrepancy is not critical given that the observed spectrum is a linear combination of multiple \APEC{} models with different temperatures and redshifts.
		We therefore simulated the expected energy spectrum of unresolved \GCG{} based on their XLF. 
		To this end we constructed a grid of $15\times15$ cells covering the relevant parameter ranges on the redshift-temperature plane. 
		For each cell on the grid we computed the \APEC{} model, which normalisation was determined according to the XLF of \GCG{},
		using the luminosity-temperature scaling relation of \citet{Giles2015}. 
		The K-corrected emission spectra of all cells were summed to obtain the theoretical spectrum of unresolved \GCG{},
		which was then fit with a single temperature \APEC{} model, similar to the observed one (\Fref{LSS_Espec_APEC}). 
		The best-fit parameters of this simulation are shown in \Fref{LSS_Espec_APEC_ChiSqr02} with the red triangle.
		As one can see, it is consistent with the best-fit values of the temperature and redshift of the observed spectrum of CXB fluctuations within $1\sigma$ contour.
		
		Based on the best-fit parameters {of the observed energy spectrum (\Fref{LSS_Espec_APEC})}, we can derive for the unresolved \GCG{} in XBOOTES
		a characteristic luminosity of $L_{\eSoft{}} = 3^{+3}_{-1}\times10^{42}\,\mathrm{erg\;s^{-1}}$ 
		and DMH mass of $M_{500}=4^{+3}_{-1}\times10^{13}\,\mathrm{M_{\sun}}\,h^{-1}$.
		For the latter we use the mass-temperature scaling relation of \citet{Lieu2016}.
		
		%----------------------------------------------------------------------------------------------------------------------
			\begin{figure}
			\begin{center} % .r p67_MSC_h_TalkPlots.pro
				\resizebox{\hsize}{!}{\includegraphics{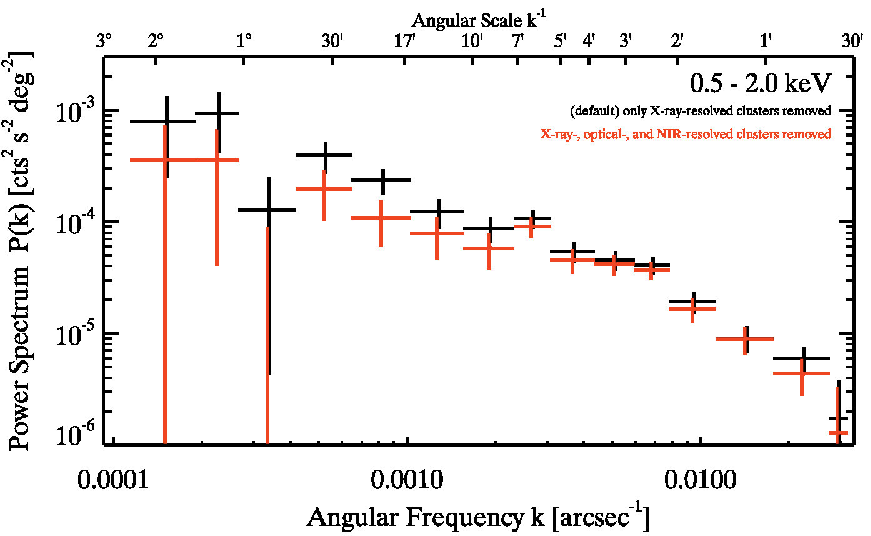}} 
				\caption{\label{f:PS_eSoft_NoPSSN_Cat} 
					Comparison of the {\PoSp{}} of the unresolved CXB, when additionally optically identified or NIR candidates of \GCG{} are removed.
					These sources were taken from catalogs listed in \Sref{s:LSS_Cat}.
					The corresponding {mask is shown on the left of \Fref{MSKs_four}}.
					}
			\end{center}
			\end{figure}
				
			\begin{figure}
			\begin{center} % .
				\includegraphics[width=0.235\textwidth]{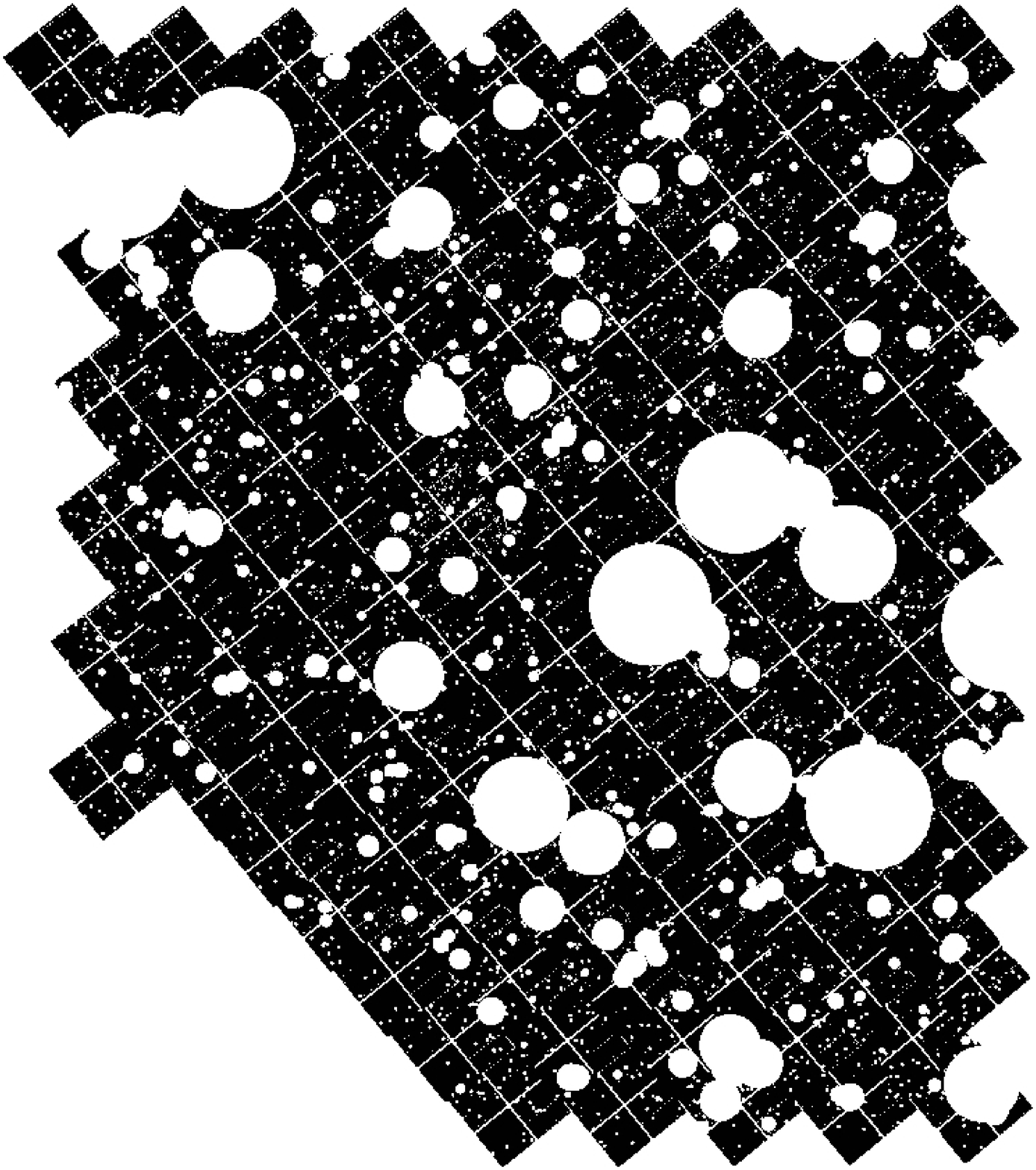}  % Xray/optical/IR groups/clusters removed
				\hfill
				\includegraphics[width=0.235\textwidth]{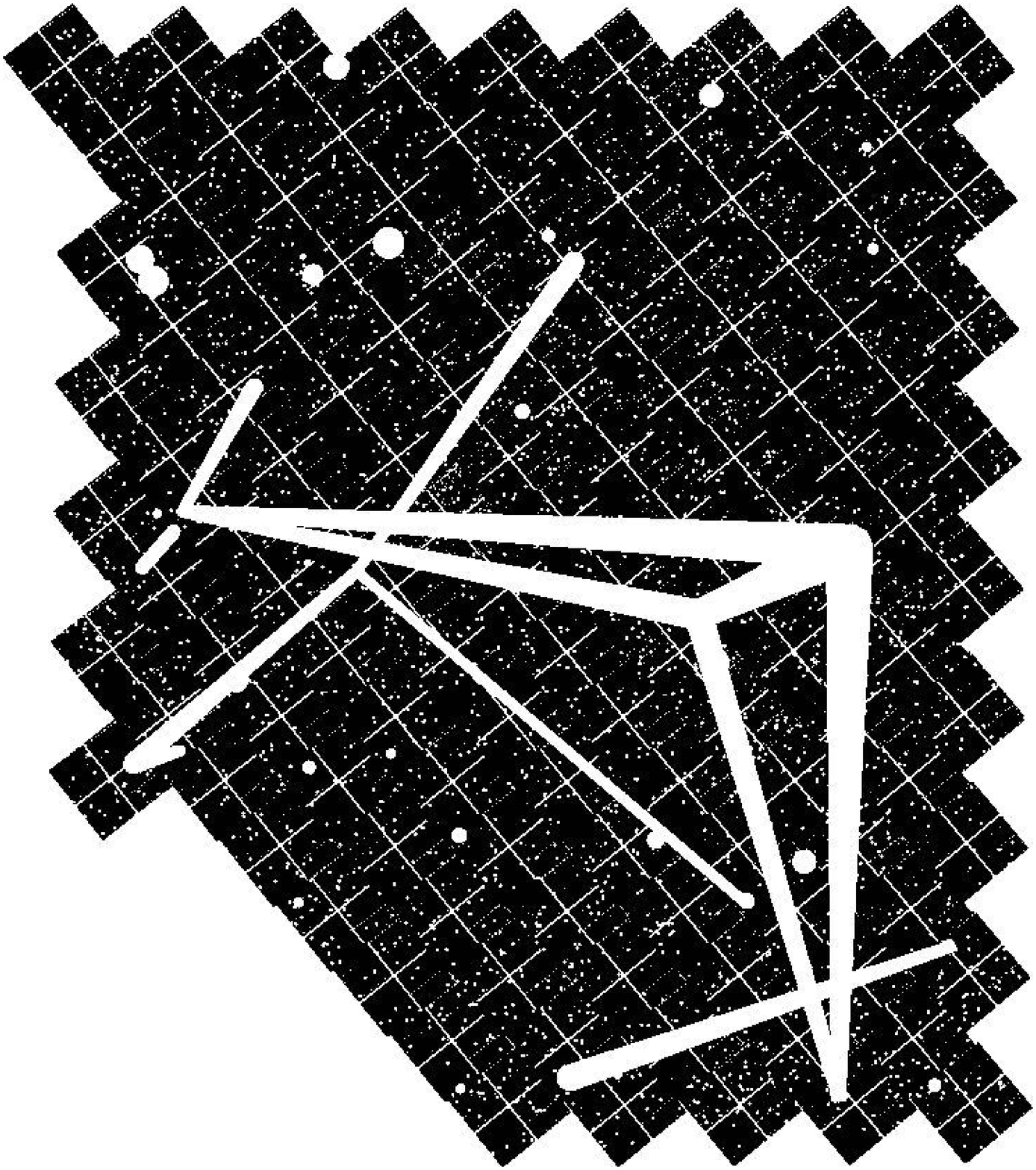}  % removes filaments of G.C which are <25Mpc apart (12 filaments)
				\caption{\label{f:MSKs_four} 
					Mosaic masks used to constrain the combined contributions of optical and NIR \GCG{} (left, see \Sref{s:LSS_Cat}) and filaments between X-ray-resolved \GCG{} (right, see \Sref{s:WHIM}). 
					They are based on the default mask (\Fref{MSK_RS10}).
					}
			\end{center}		
			\end{figure}

		%----------------------------------------------------------------------------------------------------------------------
		\subsection{Contribution of optically- and NIR-identified \GCG{}}  \label{s:LSS_Cat}
		%----------------------------------------------------------------------------------------------------------------------
		In this section we will investigate the contribution of \GCG{} identified in the XBOOTES field at other wavelengths. 
		To this end we will be excluding from the analysis the image areas around  known \GCG{} and computing the power spectrum of unresolved CXB, 
		comparing the result with our default mask. 
		We will use the following catalogs:
		\begin{enumerate}
			\item The SDSS-DR12 catalog by \citet{Tempel2017}.
			It contains \GCG{}  below redshift $z=0.2$ and brighter than  $r = 17.77$.
			About 150 of their objects  are within the area of XBOOTES survey.
			These sources are removed with a circular exclusion area with a radius equal to half of their $R_{200}$.
			\item The SDSS-DR6 catalog by \citet{Szabo2011}.
			This catalog contains \GCG{} in the redshift range of $0.1<z<0.7$ with the  lower magnitude limit of $r = 22.0$ ($>90\,\%$ completeness).   
			About 120 of their objects are within the area of XBOOTES.
			Sources are removed with a circular exclusion area with the radius equal to {half of} their $R_{200}$. 
			\item A catalog of cluster candidates of \citet{Eisenhardt2004} from the Spitzer/IRAC shallow survey (ISCS) of NDWFS. 
			The survey has the aperture-corrected 5$\sigma$ depth of $\approx19.1$ and $18.3$~mag (Vega) at 3.6 and 4.5~$\mu$m, respectively.
			It contains objects up to redshift $z=2.2$
			and about 330 of them are within the area of XBOOTES.
			We used a  circular exclusion region  of a constant radius of {$400\,\mathrm{kpc}\,h^{-1}$}, since their physical sizes are unknown.
			{This value is approximately equal to the median $R_{200}/2$ of the selected SDSS-DR6 sources, $\sim2.5$ times larger than the median $R_{200}/2$ of the selected SDSS-DR12 sources, and $\sim30\,\%$ larger than the median $R_{500}$ of 14 X-ray-resolved \GCG{} in XBOOTES obtained by \citet[Table~3]{Vajgel2014}.}
		\end{enumerate}
		
		These catalogs were used to amend our default mask, and the resulting mask%
			\footnote{
			Our definition of the physical radius would lead to very large masked out circles (radius $\sim18\arcmin-46\arcmin$) for three very nearby sources ($z<0.014$).
			To minimize the complexity of the mask effect and the reduction of the S/N, we set their angular radius to $12\arcmin$ ($\sim$fourth largest angular radius), which corresponds to $\sim100\,\mathrm{kpc}\,h^{-1}$ for the one NIR source and $\sim0.3\times\,R_{200}$ for the two SDSS-DR12 sources and reduces their combined masked area by almost an order of magnitude.
			}
		is shown on the left of \Fref{MSKs_four}. 
		The \PoSp{}%
			\footnote{
			We corrected for the stronger suppression of power on large angular scales due to the mask effect in respect to the default mask.
			It is a $\sim15\,\%$ effect on the largest considered scales.
			}
		is  shown in \Fref{PS_eSoft_NoPSSN_Cat} along with our nominal \PoSp{} of the unresolved CXB obtained with the default mask.
		One can see that,
		although the exclusion of the selected clusters does have some effect on the power spectrum,
		they can not explain the amplitude of observed fluctuations of unresolved CXB.
		This suggests that the used catalogs may be not deep and/or complete enough to account for the observed fluctuations.
		
		%---------------------------------------------------------------------
			\begin{figure}
			\begin{center} % .r p67_MSC_h_TalkPlots.pro
				\resizebox{\hsize}{!}{\includegraphics{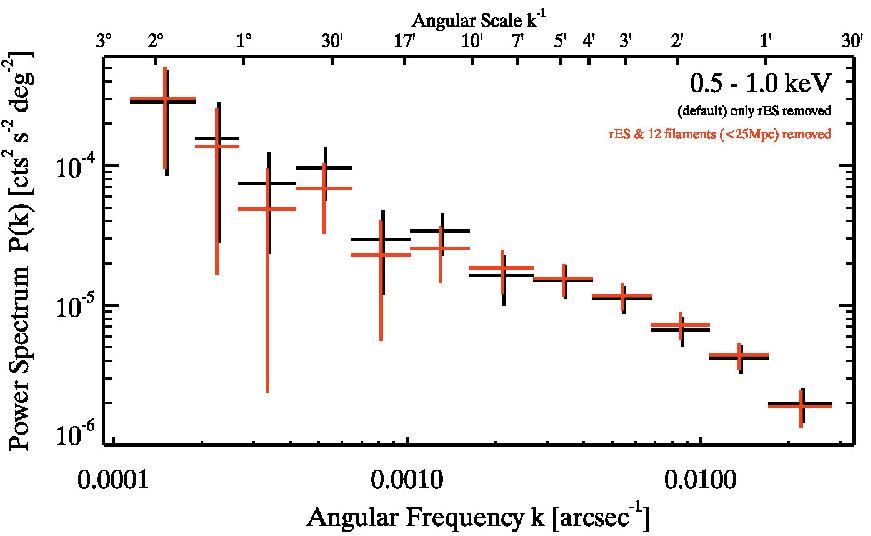}} 
				\caption{\label{f:PS_eSoft_NoPSSN_fila} 
					The \PoSp{} of the unresolved CXB in the $0.5-1.0\,\mathrm{keV}$ band, 
					where filaments between resolved \GCG{} are retain (black, default) and removed (red).
					The corresponding mask for the latter is shown in \Fref{MSKs_four} on the right.
					}
			\end{center}
			\end{figure}		
		%---------------------------------------------------------------------
		\subsection{WHIM} \label{s:WHIM}	
		%---------------------------------------------------------------------
		Due to its low temperature {\citep[$\sim10^5-10^7\,\mathrm{K}\lesssim1\,\mathrm{keV}$, see e.g. review of][]{Bregman2007}}, WHIM is not expected to make any significant contribution to the fluctuations of unresolved CXB in the most of the energy range considered in this work. 
		In agreement with this, the energy spectrum of CXB fluctuations (\Sref{s:LSS_Espec})  is adequately described by the emission  spectrum of optically thin plasma with the temperature of $1.3^{+0.4}_{-0.2}$~keV, and does not require any softer spectral component.
		This temperature is obviously too high to be associated with WHIM.
		On the other hand, our current measurement can not exclude some contribution of WHIM in the softest energy band (\eGal{}).
		
		Some hot gas may be found in the filaments between \GCG{}, as cosmological hydrodynamical simulations predict and very deep \xmm{} observations have shown for single cases \citep[e.g.][]{Cen2006,Werner2008,Roncarelli2012,Nevalainen2015}. 
		To constrain contribution  of such filaments associated with  resolved \GCG{}  we identify among the latter all pairs of clusters separated by the comoving distance smaller than $25\,\mathrm{Mpc}\,h^{-1}$. 
		This is a rather conservative upper  limit on the  filament length \citep[e.g.][]{Tempel2014b}.
		We mask out all the regions on the mosaic image connecting such pairs of clusters, in addition to all resolved sources excluded by the default mask.
		The filament regions have a trapezoidal form, which width at each end equals to the angular size of the cluster of galaxies it is connected to.
		For the angular size of a cluster of galaxies we used the diameter of its circular exclusion area of the default mask (i.e. $2\times(6\times\text{rES-size})$). 
		The resulting \PoSp{} in the $0.5-1.0\,\mathrm{keV}$ band is shown in \Fref{PS_eSoft_NoPSSN_fila} along with our standard \lssPS{} obtained with the default mask.
		As one can see from this plot, the possible filaments of gas connecting resolved \GCG{} do not make a significant contribution to the \PoSp{} of unresolved CXB.
		The same can be concluded from the comparison of the average power (\Eqref{eq:PS_Ave}, for $3\arcmin - 1^\circ$) for the lowest energy band (\eGal{}) for both masks.

	%===============================================================================================
	\section{Summary} \label{s:sum} 
	%===============================================================================================
		Surface brightness fluctuations of CXB carry unique information about faint source populations,
		which are unreachable via conventional approach based on studies of resolved sources.
		Accessing these information via angular correlation studies has become a new frontier of LSS research with X-ray surveys, 
		successfully complementing conventional studies \citep[e.g.][]{Cappelluti2012b,Cappelluti2013,Helgason2014,Mitchell2016,Kolodzig2016}.
		
		We studied fluctuations of the X-ray surface brightness in the XBOOTES field.
		With its area of $\approx 9$~deg$^2$ it is the largest contiguous \chandra{} survey which  has been observed  to the $\approx 5$~ksec depth.
		We constructed mosaic images of the entire XBOOTES field in various energy bands and, after masking out resolved sources (point-like and extended),  computed \PoSpa{} covering the range of angular scales from $\approx 2\arcsec$ to $\approx 3^\circ$.
		This  extends by more than an order of magnitude  the largest angular scales investigated in \Kol{} where stacked \PoSpa{} computed over individual \chandra{} {observations} were analyzed. 
		After subtracting the contribution of unresolved point sources (the so called \PoSoSN{}) we obtained the \PoSp{} of fluctuations of unresolved CXB.
		We also computed  \PoSp{} of the mosaic image in which only resolved point sources were masked out while all extended sources were left on the image.  
		The difference between the latter and the \PoSp{} of unresolved CXB represents the \PoSp{} of  resolved \GCG{}
		(see footnote \ref{footn:confirmed_gc} regarding identification of resolved extended  sources with \GCG{}).
		These results present the most accurate CXB fluctuation measurement to date at angular scales below $\sim3^\circ$.

		In the \PoSp{} of unresolved CXB, the non-trivial LSS signal dominates the shot noise of unresolved point sources at all angular scales above $\sim1\arcmin$.
		As it was demonstrated in  \Kol{}, this signal is mainly due to   CXB brightness fluctuations caused by unresolved clusters and groups of galaxies.

		The main results of this work can be summarized as follows:

		\begin{enumerate}

			\item There is a clear difference in shape between \PoSpa{} of unresolved CXB and resolved \GCG{}.
			While the former has an approximate power law shape with the slope of $\Gamma= 0.96\pm 0.06$ in the entire  range of angular scales, the latter is significantly steeper,  with $\Gamma=1.76\pm 0.04$, and has a clear low frequency break at 
			the angular scale of $\sim 30\arcmin$.
			The location of the low frequency break  suggests that this analysis is sensitive to the ICM structure out to $\sim3\times R_{500}$ ($\sim3\,\mathrm{Mpc}\,h^{-1}$, \Sref{s:PS_rES}). Thus, CXB fluctuations carry information about the average ICM structure  at large radii, out to the virial radius.

			\item {From the \PoSpa{} computed in a number of narrow energy bands we constructed the energy spectrum of fluctuations using the approach similar to the Fourier-frequency resolved spectroscopy proposed by \citet{Revnivtsev} to study spectral variability of X-ray binaries.
			The  energy spectra ($0.5-10.0\,\mathrm{keV}$) of  fluctuations of the unresolved CXB and of resolved \GCG{} are well described by the redshifted emission spectrum of optically thin plasma, as it should be for ICM emission.
			For fluctuations of unresolved CXB we obtained the best-fit temperature of $T=1.3^{+0.4}_{-0.2}\,\mathrm{keV}$ and the redshift  of $z=0.4^{+0.1}_{-0.2}$.
			These numbers are  consistent with theoretical expectations based on the XLF of \GCG{} and scaling relations for the parameters characterizing their X-ray emission.
			The  DMH mass corresponding to the best-fit parameters is $M_{500}=4^{+3}_{-1}\times10^{13}\,\mathrm{M_{\sun}}\,h^{-1}$ and the luminosity is $L_{\eSoft{}} = 3^{+3}_{-1}\times10^{42}\,\mathrm{erg\;s^{-1}}$.
			For resolved clusters we obtained $T=2.1^{+0.5}_{-0.3}\,\mathrm{keV}$ and $z=0.3^{+0.3}_{-0.1}$, which is in agreement with the redshift and ICM temperature of resolved \GCG{} in XBOOTES.
			As expected, fluctuations of unresolved CXB are caused by cooler (i.e. less massive) and more distant clusters and groups of galaxies.}

			\item
			Comparison with the available catalogs of \GCG{} covering the XBOOTES field suggests that they may be not deep {and/or complete} enough to account for the observed fluctuations.
			We also did not find clear evidence for contribution of WHIM to the observed fluctuations of the CXB surface brightness.

		\end{enumerate}
								
		Our results demonstrate the significant diagnostic potential of angular correlation analysis of CXB fluctuations in order to study the ICM structure in  \GCG{}.

	%----------------------------------------------------------------------------------------------------------------------
	% Acknowledgements
	%----------------------------------------------------------------------------------------------------------------------
	\section*{Acknowledgments}
		We have enjoyed helpful discussions with M. Anderson, D. Eckert, M. Krumpe, F. Zandanel, and  M. Roncarelli.
		The first author acknowledges support by China Postdoctoral Science Foundation, Grant No. 2016M590012.
		MR and RS acknowledge partial support by Russian Scientific Foundation (RNF), project 14-22-00271.
		GH acknowledges support by the Estonian Ministry of Education and Research grant IUT26-2 and EU ERDF Center of Excellence program grant TK133. 
		The scientific results reported in this article are based on data obtained from the \chandra{} Data Archive.
		This research has made use of software provided by the \chandra{} X-ray Center (CXC) in the application package CIAO. %, ChIPS, and Sherpa.

	%===========================================================
	% Appendix
	%===========================================================
	\appendix
		
	%----------------------------------------------------------------------------------------------------------------------
	\section{Photon shot noise} \label{a:PhoSN}
	%----------------------------------------------------------------------------------------------------------------------
		The \PhSN{} ($P_\mathrm{Phot.SN}$) is an additive, scale-independent component of the \PoSp{}, which arises from the fluctuation of the number of photons per beam.
		Since we are using instrumental-background-subtracted count maps $\map{C}$ (\Eref{eq:NetCtsMap}), we have to take into account the \PhSN{} of the total-count maps ($\map{C}^\mathrm{Total}$) and of the instrumental-background maps ($\map{C}^\mathrm{BKG}$, \Eref{eq:BKG}).
		Since both are uncorrelated, we can estimated their \PhSN{} separately and add them up:
			\begin{align} \label{eq:PhSNoise}
				P_\mathrm{Phot.SN}  = P_\mathrm{Phot.SN}^\mathrm{Total} + P_\mathrm{Phot.SN}^\mathrm{BKG} \text{ .} %\notag
			\end{align}
		
		For the \sPoSp{} we are using the \emph{analytical estimator} to estimate the \PhSN{} for both types of maps:
			\begin{align} \label{eq:AnaSNoise}
				P_\mathrm{Phot.SN}^\mathrm{Total}  = 
				\dfrac{1}{\Omega} \sum_j^{N} \dfrac{ C^\mathrm{Total}_j}{ E_j^2 } \text{ ,} %\notag
			\end{align}		
			\begin{align}  \label{eq:AnaSN_BKG}
				P_\mathrm{Phot.SN}^\mathrm{(BKG)} = 
				\dfrac{1}{\Omega} \sum_j^{N} \dfrac{ C_j^\mathrm{Stow} \cdot S_j^2 }{ E_j^2 } 
				\text{ .} %\notag
			\end{align}		
		For simplicity we use here a single index ($j$) for the summations over all image pixels $N$ of a 2D quantity.
		$C_j^\mathrm{Total}$, $C_j^\mathrm{Stow}$, $E_j$, and $S_j$ are pixels of the maps $\map{C}^\mathrm{Total}$, $\map{C}^\mathrm{Stow}$, $\map{E}$, and $\map{S}$, respectively.
		The stowed background map$^{\ref{StowedBKG}}$ ($\map{C}^\mathrm{Stow}$) is the same for all observation,
		while the map $\map{S}$ of each observation has the rescaling factor $S$ (\Eref{eq:BKG_res}) as a constant value.
		The analytical estimator is explained and discussed in \Kol{} (appendix~C), where we also show its derivation%
			\footnote{
				Note that the equation of $P_\mathrm{Phot.SN}^\mathrm{(BKG)}$ in appendix~C1.1 of \Kol{} is incorrect.
				The correct version is \Eqref{eq:AnaSN_BKG}.
			}.
		
		For the \mPoSp{} we are using the analytical estimator for the \PhSN{} of the total-count mosaic ($\msc{C}^\mathrm{Total}$).
		In this respect, $C_j^\mathrm{Total}$ and $E_j$ in \Eqref{eq:AnaSNoise} are pixels of the mosaics $\msc{C}^\mathrm{Total}$ and $\msc{E}$, respectively, which are constructed out of the maps $\map{C}^\mathrm{Total}$ and $\map{E}$. 
		The instrumental-background mosaic $\msc{C}^\mathrm{BKG}$ is constructed out of the maps: $(\map{C}^\mathrm{Stow} \cdot \map{S}^2)$.
		Since $\map{C}^\mathrm{Stow}$ is the same for all observation, the analytically estimator (\Eref{eq:AnaSN_BKG}) overestimates significantly the \PhSN{} (e.g. $\gtrsim10\,\%$ in the \eSoftB{}).
		Hence, for the instrumental-background mosaic we are using the \emph{high-frequency estimator},
		where we estimate the \PhSN{} from the average power of the frequency range $[k^\mathrm{(HF)}_\mathrm{min},k_\mathrm{Ny}^\mathrm{Mosaic}]$ with $k^\mathrm{(HF)}_\mathrm{min}= k_\mathrm{Ny}^\mathrm{Mosaic} \times 0.80 \approx 0.025\InvArcSec \approx ( 39\arcsec )^{-1} $.
		This is possible because for the chosen frequency range the photon-shot-noise-subtracted \PoSp{} of the stowed background map ($\map{C}^\mathrm{Stow}$) is about two orders of magnitude smaller than the \PhSN{} itself. 
		The high-frequency estimator is explained and discussed in detail in \Kol{} (appendix~C). 
		
		For the \mPoSp{} the \PhSN{} of the instrumental-background $P_\mathrm{Phot.SN}^\mathrm{BKG}$ contributes less than $1\,\%$ to the total \PhSN{} (\Eref{eq:PhSNoise}). 
		Hence, given the shape and amplitude of our \cxbPS{} (\Fref{PS_eSoft}) only small angular scales below $\sim1\arcmin$ of the \mPoSp{} are affected by potential inaccuracies of the estimate of $P_\mathrm{Phot.SN}^\mathrm{(BKG)}$.
		In this respect, one should take in mind that our combined \PoSp{} (\Sref{s:combinedPS}) only takes the \mPoSp{} above $\sim1\arcmin$ (below $k_\mathrm{C} = k_\mathrm{Ny}^\mathrm{Mosaic} / 2 \approx 0.016\InvArcSec$) into account, which is a precaution in order to avoid the smallest angular scales of the \mPoSp{} (\Aref{a:PSs}).
		
		Note, that all shown \PoSpa{} in this work are already subtracted by the \PhSN{}.
	
	%----------------------------------------------------------------------------------------------------------------------
		\begin{figure}
		\begin{center} % .
			% set f="PSF_Regions_on4238"
			% convert ${f}.jpg eps2:${f}.ps
			% left bottom right top [trim = 90mm 20mm 90mm 10mm, clip]
			\resizebox{\hsize}{!}{\includegraphics{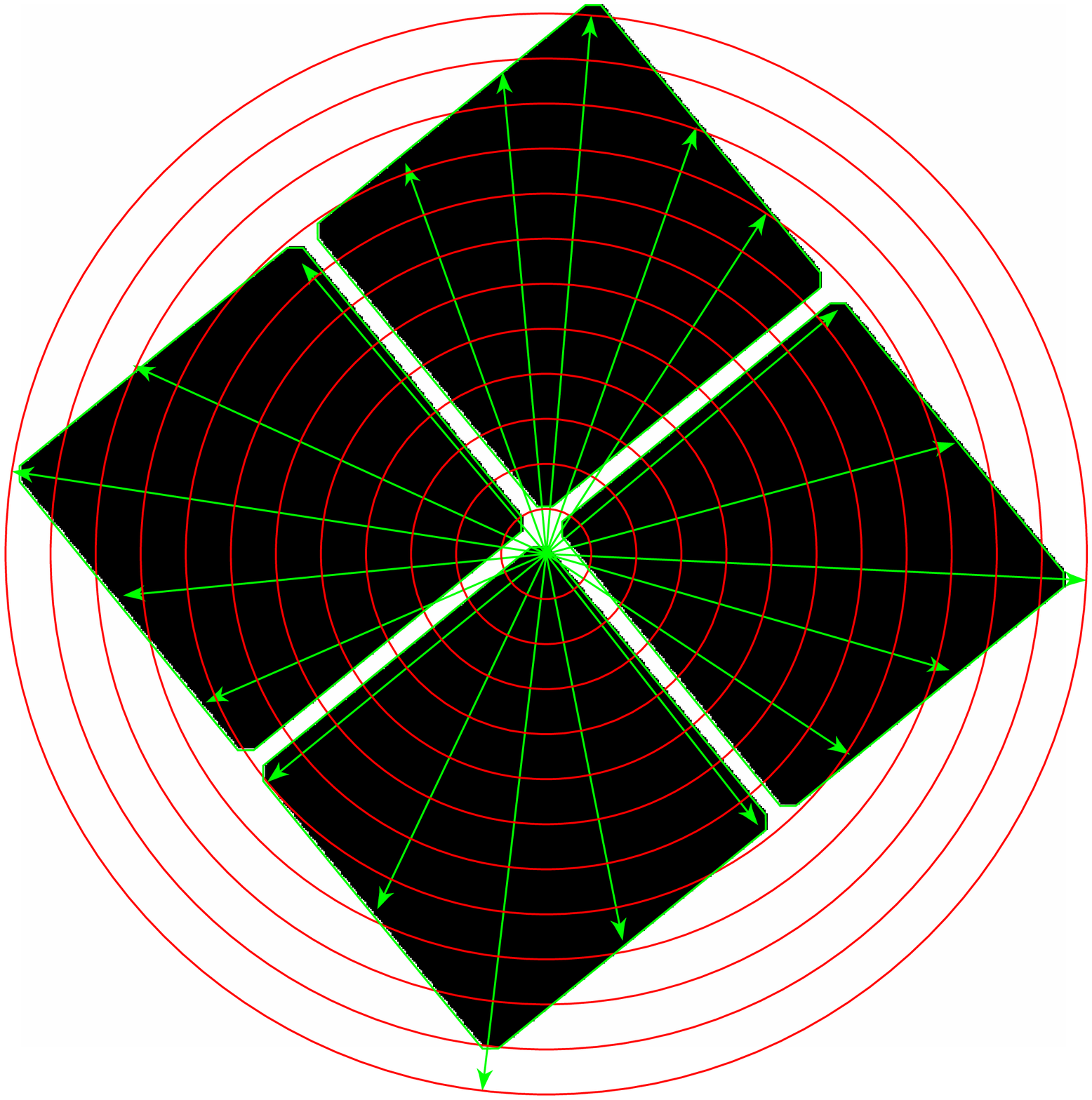}} 
			\caption{\label{f:PSF_pos} 
				Sampling of \acisi{}'s FOV for the PSF-smearing model.
				Green arrows show the 20 azimuthal angles and red circles show the 13 offset angles ($\theta=0\arcmin-12\arcmin$, in $1\arcmin$-steps).
				PSF simulations were performed at positions, where green and red lines intersect within the black areas.
				See \Aref{a:PSFsmear} for details.
				}
		\end{center}		
		\end{figure}
		
		\begin{figure}
		\begin{center} % .
			\resizebox{\hsize}{!}{\includegraphics{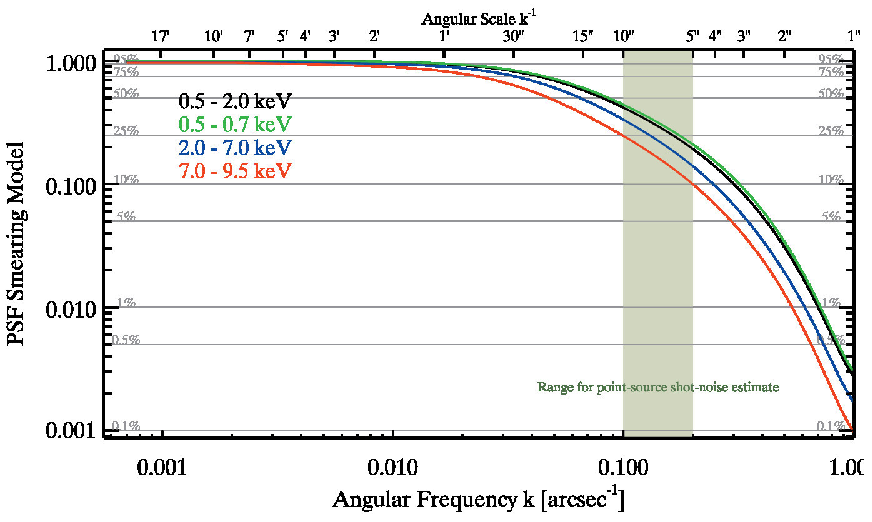}} 
			\caption{\label{f:PSF_PoSp}
				Our PSF-smearing model for selected energy bands.
				See \Aref{a:PSFsmear} for details.
				}
		\end{center}		
		\end{figure}	
	%----------------------------------------------------------------------------------------------------------------------
	\section{PSF-smearing model} \label{a:PSFsmear}
	%----------------------------------------------------------------------------------------------------------------------
		Given our change in the data preparation and processing in comparison to \Kol{} (\Sref{s:Change}), we also update our PSF-smearing model $W_\mathrm{PSF}(k)$ (for previous model see appendix~B of \Kol{}), which is now computed for each energy band individually.
		For the PSF simulation we are using the CIAO tool \texttt{simulate\_psf} in combination with the MARX software package%
				\footnote{\url{http://space.mit.edu/ASC/MARX}}
			(v5.3.1).
		We adjust the count rate in each band to achieve an on-axis pile-up probability below $\sim1\,\%$.
		To still obtain a high S/N, we use an exposure time of $\sim1.3$~Ms, which results in more than $\sim2500$ counts per PSF simulation.
		We sample \acisi{}'s FOV with 20 azimuthal angles and 13 offset angles ($\theta=0\arcmin-12\arcmin$, in $1\arcmin$-steps).
		This results in 173 unique PSF positions within the FOV mask, as shown in \Fref{PSF_pos}.
		We compute the \PoSpa{} of all PSF simulations and first average them over all azimuthal angles per offset angle before we compute the weighted average over all offset angles.
		For the weights we use the surface area times the average exposure time of the annulus ($1\arcmin$ wide) of each offset angle.
		This weighting is designed to also take the vignetting into account, although it appears to be almost neglectable effect.
		The resulting FOV-averaged PSF \PoSp{} $W_\mathrm{PSF}(k)$, alias our PSF-smearing model, is shown in \Fref{PSF_PoSp} for selected energy bands.
		
		We can see from \Fref{PSF_PoSp} that the PSF-smearing is only important for angular scales below $\sim1\arcmin$ and its impact increase with energy as expected given the energy dependence of \chandra{}'s PSF\footnote{\url{http://cxc.harvard.edu/proposer/POG/html/chap4.html#tth_sEc4.2.3}}.

	%----------------------------------------------------------------------------------------------------------------------
	\section{Systematic effects} \label{a:Sys}
	%----------------------------------------------------------------------------------------------------------------------
			
		Below we discuss several systematic effects of our measurement of the CXB surface brightness fluctuations.
		Here, we focus primarily on the \mPoSp{} since we already studied extensively the systematic effects of the \sPoSp{} in \Kol{} (appendix~D).

		%-----------------------------------------------------------
		% Figure:  Mass Effect
		%----------------------------------------------------------- 
			\begin{figure}
			\begin{center}
				\resizebox{\hsize}{!}{\includegraphics{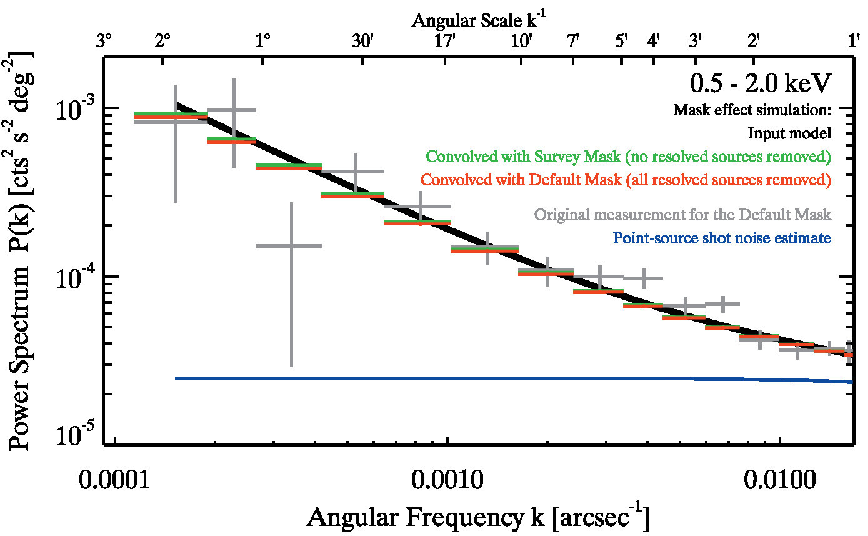}} 
				\resizebox{\hsize}{!}{\includegraphics{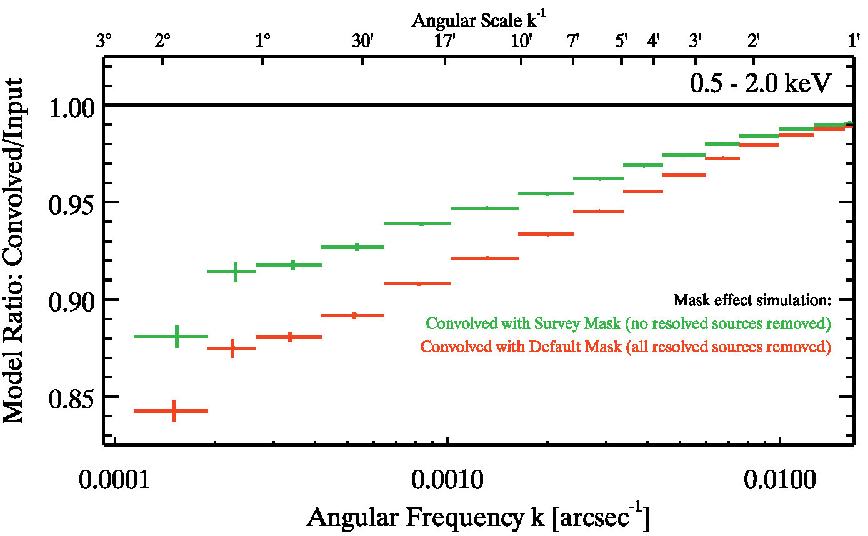}} 
				\caption{\label{f:MaskEff}
				The impact of the mask effect on the \mPoSp{}. 
				\emph{Top:}
					The convolved mosaic \PoSpa{} after the default mask (red) or the survey mask (green) was applied to a simulated image, which is based on the input model (black curve).
					Also show is the \mPoSp{} of the original XBOOTES observations for the  default mask (gray).
				\emph{Bottom:} 
					The ratio of the convolved mosaic \PoSpa{} to the input model.
				See \Aref{a:MCS} for details.
				}
			\end{center}
			\end{figure}
		%-----------------------------------------------------------
			
		\subsection{Mask effect} \label{a:MCS}
		%-----------------------------------------------------------
		The impact of the mask effect on the \sPoSp{} is shown in \Kol{} (appendix~D1).
		We use the same procedure describe there to test the impact on the \mPoSp{} by the default mask (\Fref{MSK_RS10}) and the \emph{survey mask}, which is the mosaic of all FOV masks (\Sref{s:ETM_MSK}).
		The input model follows \Eqref{eq:PS_Meas} and consists of the best-fit powerlaw model of the \lssPS{} for the default mask (left panel of \Fref{PS_eSoft_NoPSSN}) plus the measured \PoSoSN{} (\Sref{s:PoSoSN}), which are both multiplied by our PSF-smearing model (\Aref{a:PSFsmear}).
		We use 5000 iterations for our mask effect simulation and the resulting \emph{convolved} mosaic \PoSpa{} are shown in \Fref{MaskEff}.
		
		One can see that due the mask effect the \mPoSp{} is suppressed by less than $\sim20\,\%$ at the lowest considered frequency bin and at angular scales below $\sim30\arcmin$ it is suppressed by less than $\sim10\,\%$.
		\Fref{MaskEff} also shows that the survey geometry of XBOOTES, represented by the survey mask (green), causes the largest suppression, while the additional removal of resolved sources, included in the default mask (red), increases the suppression only by less than $\sim5\,\%$ in respect to the survey geometry.
		In any case, we can see in the top panel of \Fref{MaskEff} that the suppression is much smaller than the statistical uncertainty of our measurement (gray), which makes the mask effect an almost negligible systematic effect (also see \Aref{a:PSs}).
		This is consistent with the conclusion for the \sPoSp{} shown in \Kol{} (appendix~D1).
		
		%-----------------------------------------------------------
		% Figure: Instrumental background
		%----------------------------------------------------------- 
			\begin{figure}
			\begin{center}
				\resizebox{\hsize}{!}{\includegraphics{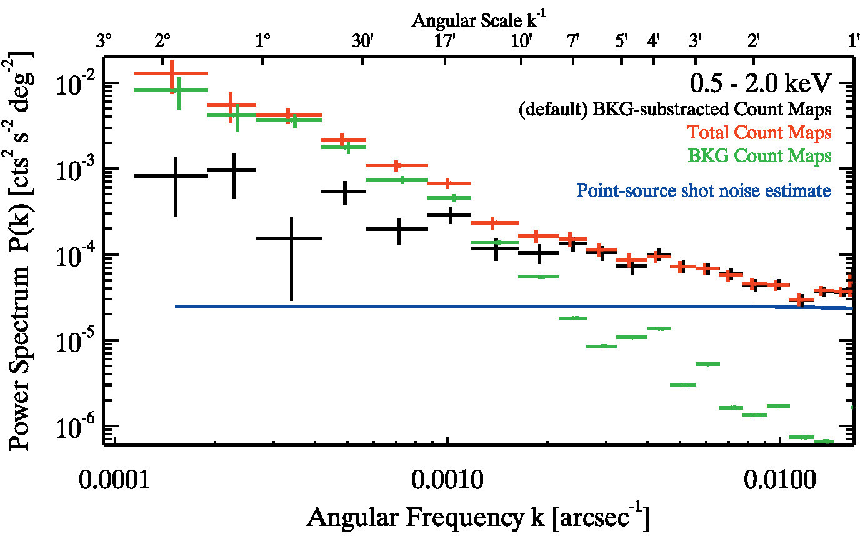}} 
				\caption{\label{f:PS_BKG}%
					%\At{Text}
					Comparison of the mosaic \PoSpa{} in the \eSoftB{} for the default mask based on three different count maps.
					In black we show our default \PoSp{}, where we use the instrumental-background-subtracted count maps $\map{C}$ (\Eref{eq:NetCtsMap}).
					In red we show the \PoSp{} for the total-count maps $\map{C}^\mathrm{Total}$, where the instrumental background was not subtracted.
					In green we show the \PoSp{} for the instrumental background maps $\map{C}^\mathrm{BKG}$ (\Eref{eq:BKG}), which are all based on the same stowed background map $\map{C}^\mathrm{Stow}$ normalized for each observation with a scaling factor $S$ (\Eref{eq:BKG_res}).
					See \Sref{s:C_maps} and \Aref{a:Bkg} for details.
				}	
			\end{center}
			\end{figure}
			
			\begin{figure}
			\begin{center}
				\resizebox{\hsize}{!}{\includegraphics{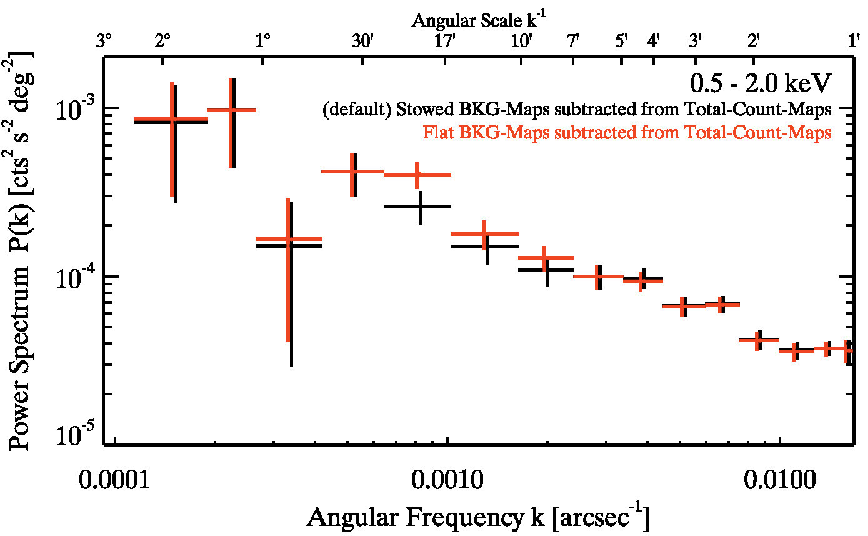}} 
				\caption{\label{f:PS_BKG02}%
					Comparison of the mosaic \PoSpa{} in the \eSoftB{} for the default mask
					based on two different methods of subtracting the instrumental background from the total-count maps ($\map{C}^\mathrm{Total}$).
					In black we show our default method, where the scaled stowed background maps $\map{C}^\mathrm{BKG}$ (\Eref{eq:BKG}) are used, and in red we show an alternative method, where flat background maps with the average surface brightness of $\map{C}^\mathrm{BKG}$ of each observation as a constant value are used.
					See \Aref{a:Bkg} for details.
				}	
			\end{center}
			\end{figure}			
		%-----------------------------------------------------------
		\subsection{Instrumental background} \label{a:Bkg}		
		%-----------------------------------------------------------
		%\At{: run for RS10, eSoft, FOV92, OL00: DT06}
		
		In \Fref{PS_BKG} we compare the \PoSpa{} of two mosaics, which are constructed either with instrumental-background-subtracted count maps $\map{C}$ (black, default, \Eref{eq:NetCtsMap}) or 
		with total-count maps $\map{C}^\mathrm{Total}$ (red, \Sref{s:C_maps}), for the \eSoftB{} for the default mask.
		Additionally, we show the \PoSp{} of the mosaic, which is constructed out of the instrumental background maps $\map{C}^\mathrm{BKG}$ (\Eref{eq:BKG}).
		We can see in \Fref{PS_BKG} that when one uses total-count maps the \PoSp{} (red) above angular scales of $\sim10\arcmin$ is dominated by fluctuations from the instrumental background (green).
		Such additional fluctuations are caused by the strong variation of the quiescent instrumental background between adjacent observations.
		Fortunately, such instrumental fluctuations can be removed by using instrumental-background-subtracted count maps.
		
		In \Fref{PS_BKG02} we present the \PoSpa{}  of two mosaics, where we used two different methods of subtracting the instrumental background from the total-count maps ($\map{C}^\mathrm{Total}$).
		Our default method is shown in black, where we use the scaled stowed background maps $\map{C}^\mathrm{BKG}$ (\Eref{eq:BKG}) for the subtraction.
		An alternative method is shown in red, where we use \emph{flat background maps} with the average surface brightness of $\map{C}^\mathrm{BKG}$ of each observation as a constant value.
		The latter method has the advantage that it is much simpler to compute and that it does not increases the overall \PhSN{} ($P_\mathrm{Phot.SN}^\mathrm{(BKG)}=0$) in comparison to our default method (\Aref{a:PhoSN}).
		We can see in \Fref{PS_BKG02} that the alternative method works almost as good as our default method but for angular scales around $\sim7\arcmin - \sim20\arcmin$ is produces slightly higher power.
		This deviation arises from inhomogeneities of the instrumental background within the FOV, which are discussed in \Kol{} (appendix~D2).
		They can only be corrected properly with the use of the scaled stowed background maps.
		However, the alternative method still appears sufficient, when one only likes to study fluctuations for angular scales at least twice as large as the FOV.
		
		Note, that if one does not include those observations with a particular high instrumental background 
		when constructing the mosaic, than the instrumental-background-subtraction would not be necessary at the given S/N of the \PoSp{} in the \eSoftB{}.
		
		For energy bands above $\sim3$~keV the instrumental fluctuations still dominate the \PoSp{} on large angular scales (see left panel of \Fref{LSS_Espec_APEC}), although instrumental-background-subtracted count maps are used.
		This arises from the fact that the effective area$^{\ref{acisi_area}}$ of \acisi{} is significantly smaller at these energies in comparison to the \eSoftB{}, which results in a much smaller fraction of source counts in respect to instrumental background counts.
		In the \eSoftB{} the fraction is $52\pm1\,\%$, while in the \eHard{} the fraction is only $8.6\pm0.3\,\%$.
		In the extreme regime, where less than one out of ten detected counts is an actual source count, our instrumental-background-subtraction method is apparently not accurate enough to remove instrumental fluctuations sufficiently well on large angular scales ($\lesssim17\arcmin$).
		Fortunately, we can account for this in our energy spectrum analysis of the \lssPS{} by including an instrumental background model in our a spectral model (\Sref{s:LSS_Espec}).

		We have already shown in \Kol{} (appendix~D2) that for the \sPoSp{} we can neglected instrumental fluctuations for angular scales within \acisi{}'s FOV ($\lesssim17\arcmin$) in the \eSoftB{}.
		We tested that this is also true with the data processing of this work (\Sref{s:DataProc}).

		%-----------------------------------------------------------
		% Figure: randomized observations
		%----------------------------------------------------------- 
			\begin{figure}
			\begin{center}
				\resizebox{\hsize}{!}{\includegraphics{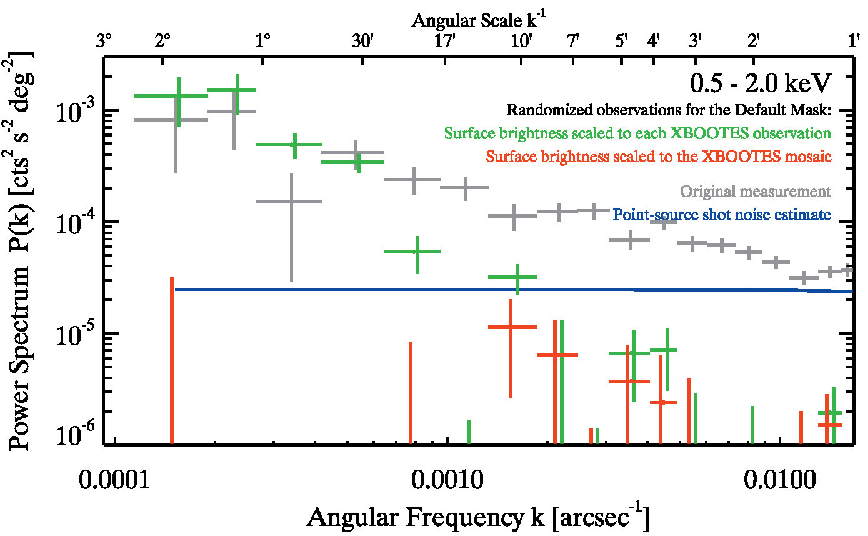}} 
				\resizebox{\hsize}{!}{\includegraphics{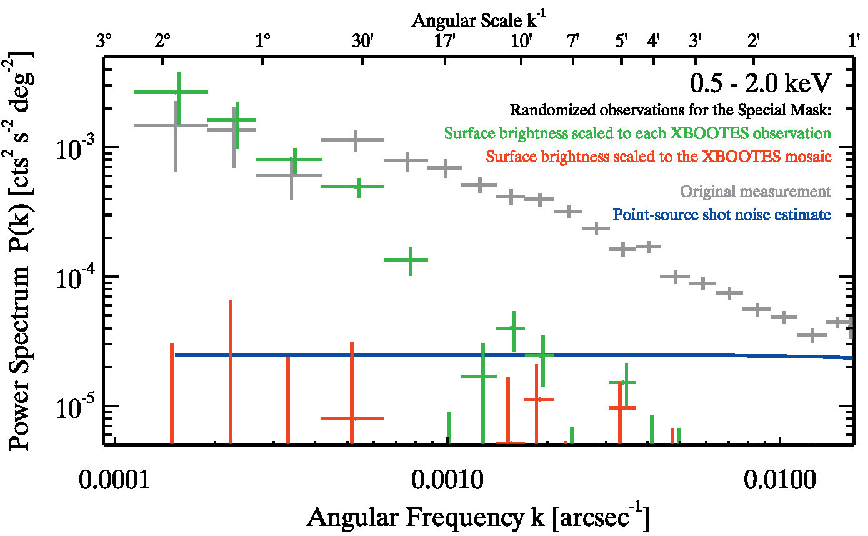}} 
				\caption{\label{f:PS_rnd}%
					Comparison of the \mPoSp{} for the original XBOOTES observations (gray), and randomized observations,
					where the  surface brightness equals the surface brightness of the original observation (green) or of the original mosaic (red).
					\emph{Top:} Default mask. % (\Fref{MSK_RS10}).
					\emph{Bottom:} Special mask.% (\Fref{MSK_RS09}).
					See \Aref{a:RND} for details.
				}	
			\end{center}
			\end{figure}
		\subsection{Test with randomized observations} \label{a:RND}
		%-----------------------------------------------------------
		%\At{: run for DT07, RS10, eSoft, FOV92, OL00: RND07 and RND05}
		To further test for instrumental fluctuations in the \mPoSp{} we also compute the \PoSpa{} of mosaics constructed with randomized observations.
		These observations are at the same sky position as the original XBOOTES observations but they only contain counts with random sky coordinates, while the total number of counts is adjusted to a certain surface brightness.
		Hence, each randomized observation only contains Poisson noise.
		We create them for each mask and energy band separately.
		The randomization smooths out any fluctuations on angular scales below \acisi{}'s FOV ($\lesssim17$), 
		which is acceptable since in this experiment we are interested on fluctuations on larger angular scales.
		We compute the \mPoSp{} for two cases:
		(a) the surface brightness of a single randomized observation equals the surface brightness of the XBOOTES observation at the same sky position,
		(b) the surface brightness of all randomized observations equals the surface brightness of the XBOOTES mosaic (\Sref{s:MSC_flux}).
		We would expect 
		for (a) that the resulting \mPoSp{} is in agreement with the original one on the largest angular scales ($\gtrsim1^\circ$),
		and for (b) that the resulting \mPoSp{} does not contain a signal at all (i.e. it is in agreement with the \PhSN{}),
		for the case that the original \mPoSp{} does not contain any additional instrumental signal due to its construction (\Sref{MSC:const}).
		In \Fref{PS_rnd} we compare all three mosaic \PoSpa{} for the default mask (top panel) and special mask (bottom panel) and we can see that they indeed agree with our expectations.
		
		%-----------------------------------------------------------
		% Figure: Overlap
		%----------------------------------------------------------- 
			\begin{figure}
			\begin{center}
				\resizebox{\hsize}{!}{\includegraphics{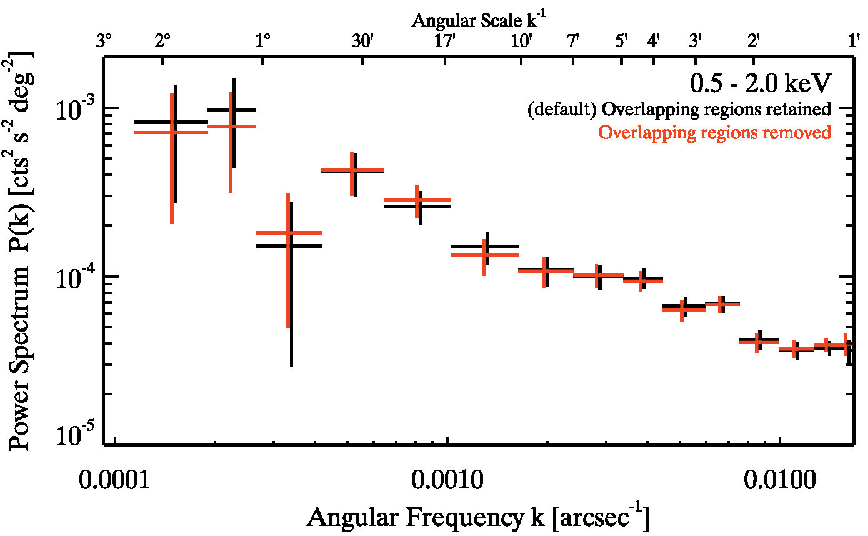}} 
				\caption{\label{f:PS_OL}%
					Comparison of the mosaic \PoSpa{} in the \eSoftB{} for the default mask, 
					where overlapping regions are retained (black, default) and removed (red).
					See \Aref{a:OL} for details.
				}	
			\end{center}
			\end{figure}
		%-----------------------------------------------------------
		\subsection{Overlap} \label{a:OL}
		%-----------------------------------------------------------
		%\At{: run for DT07, RS10, eSoft, FOV92: OL01}
		Overlapping regions between different observations account only for $\sim5\,\%$ of the total surface area of our constructed mosaic (\Sref{MSC:const}).
		To nevertheless make sure that those regions do not create any additional instrumental fluctuations, we compare in \Fref{PS_OL} the mosaic \PoSpa{} for two cases, where overlapping regions are retained (black, default) and removed (red).
		Since both \PoSpa{} agree with each other, it shows that including overlapping regions does not significantly change the \PoSp{} and it further suggests that those regions are properly taken into account when constructing the mosaic.

		%-----------------------------------------------------------
		% Figure: Circular exclusion area of \rES{}
		%----------------------------------------------------------- 
			\begin{figure}
			\begin{center} % .r p67_MSC_h_TalkPlots.pro
				\resizebox{\hsize}{!}{\includegraphics{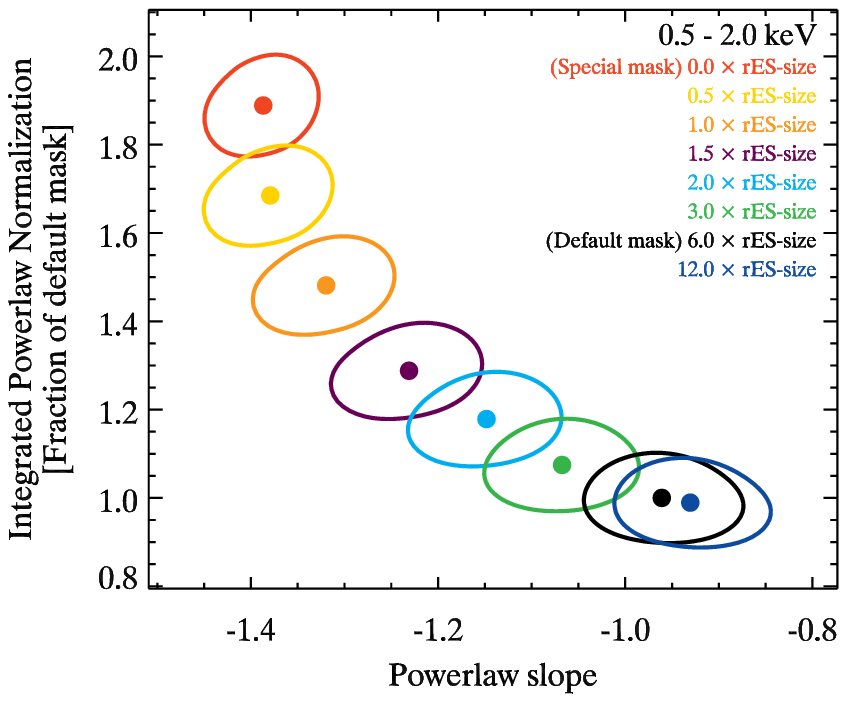}} 
				%\hfill
				\resizebox{\hsize}{!}{\includegraphics{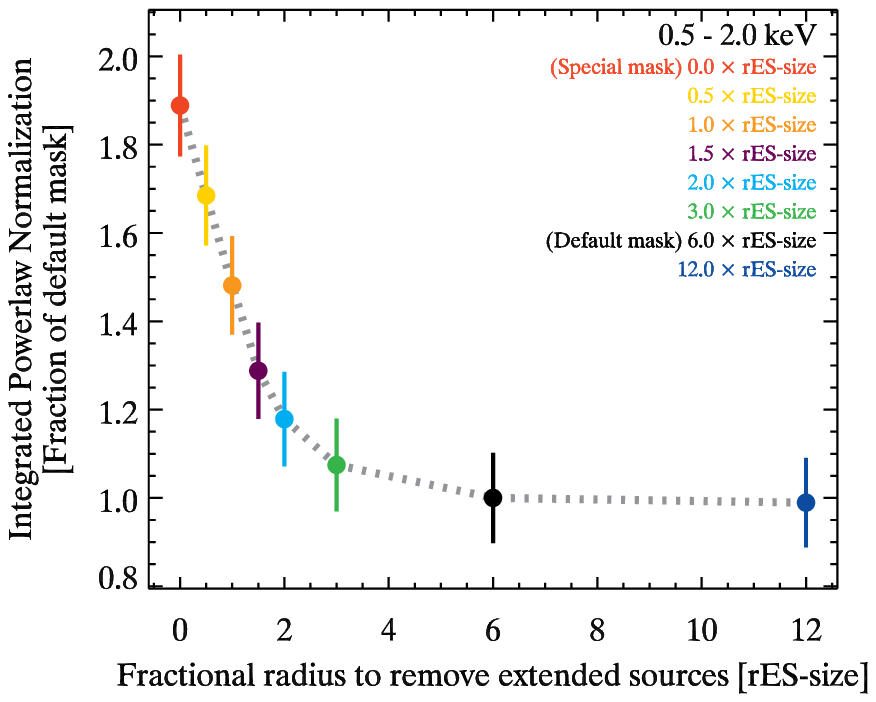}} 
				\caption{\label{f:Esize_Fit} 
					Best-fit parameters of the powerlaw fit of the \lssPS{} (\Sref{s:LSS_PL}) for different fractional radii of the circular exclusion area of \rES{}.
					Solid ellipses and lines represent one standard deviation of a two parameter fit.
					The gray dotted curve on the bottom panel is just for visualization purposes.
					The powerlaw normalization is obtained by integrating the model over the angular scale range $10\arcsec - 20\arcmin$ (Units: $\mathrm{counts^2\,s^{-2}\,deg^{-3}}$).
					See \Aref{a:LSS_rES_frac} for details.
				}
			\end{center}
			\end{figure}		
		%---------------------------------------------------------------------		
		\subsection{Circular exclusion area of \rES{}} \label{a:LSS_rES_frac}	
		%---------------------------------------------------------------------
		%\Ann{As of 02.06.2017. Move from \Sref{s:PS_rES}.}
		
		Here, we test how the \lssPS{} (\Sref{s:LSS_PL}) changes, when we gradually increase the radius of the circular exclusion area of \rES{}.
		We test eight different cases, where the radius is between $0.0$ and $12.0$ times the size of \rES{} (short: rES-size), which was determined by \Ken{} (Table~1).
		The case of $0.0\times\text{rES-size}$ represents our special mask (\Fref{MSK_RS09}), where all \rES{} are retained, while $6.0\times\text{rES-size}$ represents our default mask (\Fref{MSK_RS10}).
		The best-fit parameters of the powerlaw fit for the \lssPS{} are shown in \Fref{Esize_Fit}.
		They suggests that for our default mask, we are able to remove sufficiently well the correlation signal of resolved \GCG{} in comparison to the correlation signal of unresolved ones.
		They also suggests that the \lssPS{} is sensitive to the structure of the \GCG{}, alias the surface-brightness profile of the ICM.
		However, proper modeling is necessary in order to substantiate this quantitatively (e.g. \PapIII{}).
		
		%-----------------------------------------------------------
		% Tables: 
		%-----------------------------------------------------------
		\begin{table*}
		\caption{Main characteristics of the combined, mosaic, and stacked \PoSpa{}.}
		\label{tab:PSs}
		\begin{center}     % used for centering Table
		\begin{tabular}{l c c c}	
			\hline
			\hline
			Name	&  Angular scales & Angular frequencies	            & Based on \\
				&     $\theta$    & $\log_{10}(k[\mathrm{arcsec^{-1}}])$	&  \\
			\hline
			Combined \PoSp{} $P(k)$ & $[\sim1\arcsec,\sim3^\circ]$   & $[-4.0,-0.0]$ & $P(k) = P_S[\sim1\arcsec,k_\mathrm{C}] + P_\mathrm{M}[k_\mathrm{C},\sim3^\circ]$ \\  %  $P_\mathrm{C}(k)$
			Mosaic  \PoSp{} $P_\mathrm{M}(k)$ & $[\sim32\arcsec,\sim3^\circ]$  & $[-4.0,-1.5]$ & fluctuation mosaic ($\delta\msc{F}$), image-pixel-binning $b = 32$ \\    
			Stacked \PoSp{} $P_S(k)$  & $[\sim1\arcsec,\sim17\arcmin]$ & $[-3.0,-0.0]$ & fluctuation maps ($\delta\map{F}$), image-pixel-binning $b = 1$    \\ 
			\hline
		\end{tabular}
		\end{center}
			The lower limit in angular scales is defined by the Nyquist-Frequency $k_\mathrm{Ny} = ( 2\,b\,\Delta p )^{-1}$,
			where $\Delta p$ is \acisi{}'s chip-pixel-size ($0.492\arcsec$) and $b$ is the image-pixel-binning of the fluctuation mosaic ($b=32$) or fluctuation maps ($b=1$).
			The upper limit in angular scales for the \sPoSp{} is defined by \acisi{}'s FOV,
			while the upper limit for the \mPoSp{} is defined by the geometry of the XBOOTES survey.
			The combine frequency is $k_\mathrm{C} = k_\mathrm{Ny}^\mathrm{Mosaic} / 2 \approx ( 63\arcsec )^{-1} \approx 0.016\InvArcSec $.
			Also see \Sref{s:Def} and \Aref{a:PSs}.
		\end{table*}
		%-----------------------------------------------------------
		% Figure: Stacked, mosaic, and combined
		%----------------------------------------------------------- 
			\begin{figure}
			\begin{center}
				\resizebox{\hsize}{!}{\includegraphics{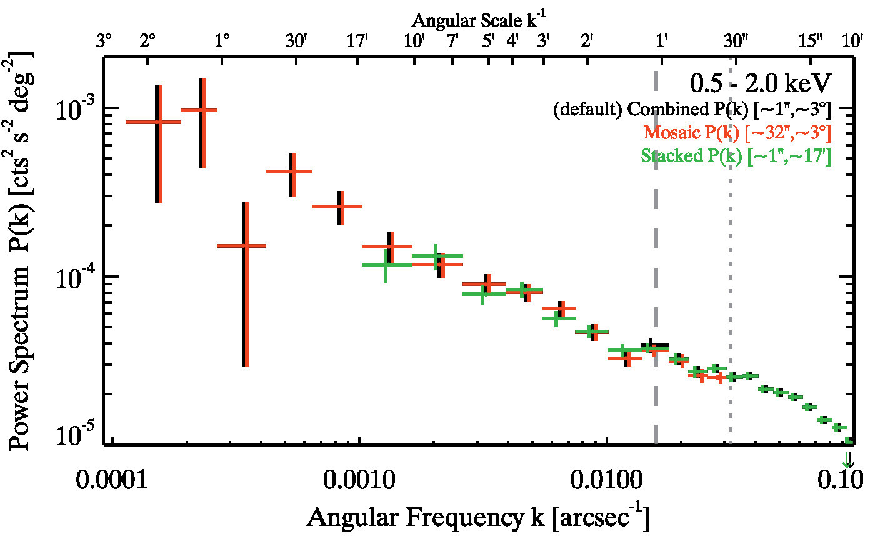}} 
				\resizebox{\hsize}{!}{\includegraphics{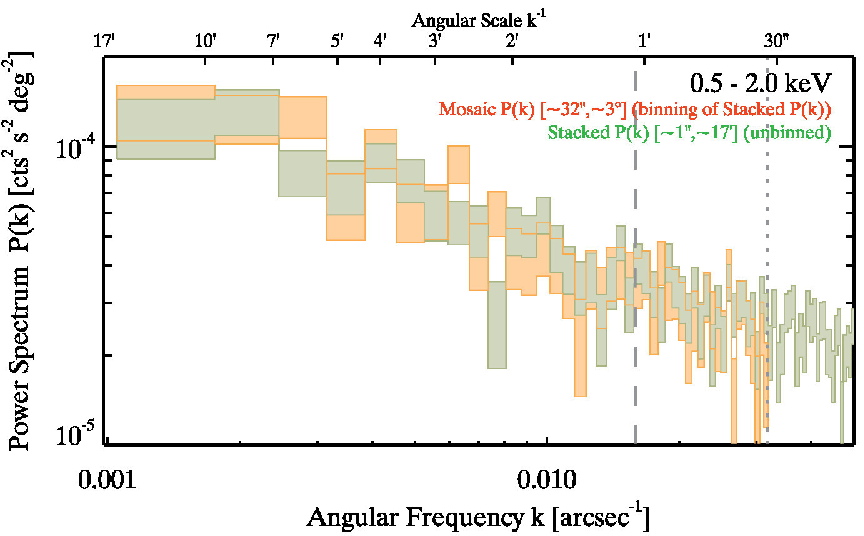}} 
				\caption{\label{f:PS_combined}%
					Comparison of the combined (black, default), mosaic (red) and stacked (green) \PoSpa{} in the \eSoftB{} for the default mask.
					Also shown are the combine frequency ($k_\mathrm{C}$) as gray dashed vertical line
					and the Nyquist frequency for the mosaic \PoSp{}  ($k_\mathrm{Ny}^\mathrm{Mosaic}$) as gray dotted vertical line.
					\emph{Top:} All \PoSpa{} are adaptively binned (default).
					\emph{Bottom:} The \sPoSp{} is unbinned, while the binning of the \mPoSp{} matches the unbinned \sPoSp{}.
					See \Sref{s:combinedPS} and \Aref{a:PSs} for details.
				}	
			\end{center}
			\end{figure}
		%-----------------------------------------------------------
		\subsection{Comparison of mosaic and stacked \PoSpa} \label{a:PSs}
		%-----------------------------------------------------------
		In \Fref{PS_combined} we directly compare the mosaic (red) and stacked (green) \PoSpa{} defined in \Sref{s:Def} and summarized in \Tref{tab:PSs}.
		We can see that the stacked and mosaic \PoSpa{} agree rather well, although it is not a perfect match.
		We notice some small modulations, which can be seen as our systematic uncertainties between the stacked and mosaic \PoSpa{}.
		They arise mainly from the (uncorrected) mask effect,
		which correlates adjacent Fourier frequencies 
		and suppresses the \PoSp{} on large angular scales (for the latter see \Aref{a:MCS} for the mosaic \PoSp{} and appendix~D.1 of \Kol{} for the stacked \PoSp{}).
		Since the dimensions of the masks used for the mosaic and stacked \PoSpa{} are an order of magnitude different ($\approx3.66^\circ$ and $\approx0.40^\circ$ side length, respectively, \Sref{MSC:const}),
		the impact of the mask effect onto the \PoSp{} is also different for a given angular scale.
		To demonstrate that stacked and mosaic \PoSpa{} are essentially fluctuating around a \emph{true} (i.e. mask effect corrected) \PoSp{},
		we also show in the bottom panel of \Fref{PS_combined} the unbinned \sPoSp{} in comparison to the \mPoSp{}, which binning matches the unbinned \sPoSp{}.
		
		We set the combine frequency ($k_\mathrm{C}$) to be two times smaller than the Nyquist-Frequency of the \mPoSp{}
		($k_\mathrm{C} = k_\mathrm{Ny}^\mathrm{Mosaic} / 2 \approx 0.016\InvArcSec \approx ( 63\arcsec )^{-1} $)
		as a precaution in order to avoid possible systematic uncertainties in estimating the \PhSN{} of the \mPoSp{} (\Aref{a:PhoSN}),
		and possible numerical inaccuracies of our discrete Fourier transform (section~4.1 of \Kol{}), which become important very close to the Nyquist-Frequency \citep[e.g.][]{Jing2005}.

		%-----------------------------------------------------------			%-----------------------------------------------------------
		% Figure: Stacked, mosaic, and combined
		%----------------------------------------------------------- 
		
			\begin{figure}
			\begin{center}
				\resizebox{\hsize}{!}{\includegraphics{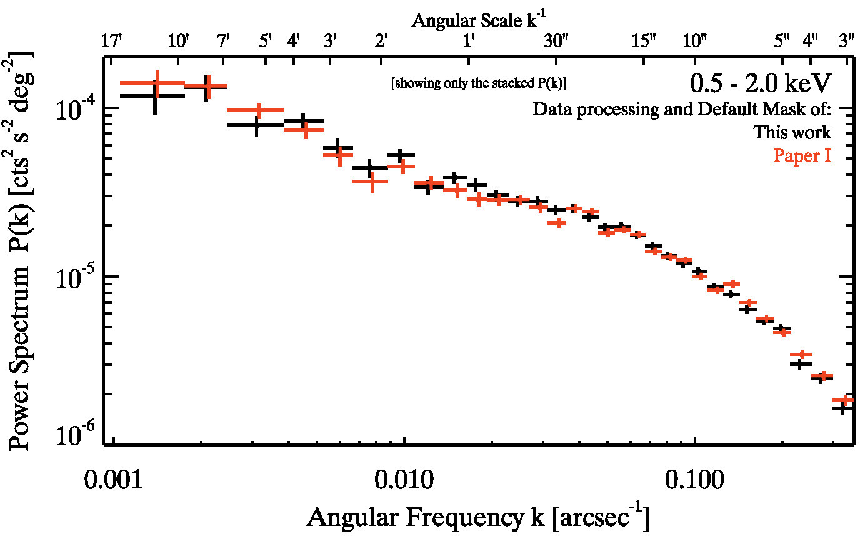}} 
				\caption{\label{f:PS_proc}%
					Comparison of the stacked \PoSpa{} in the \eSoftB{} for the data processing of this work (black) and of \Kol{} (red) for the default mask.
					See \Sref{s:Change} and \Aref{a:Data} for details.
				}
			\end{center}
			\end{figure}
			
			\begin{figure}
			\begin{center}
				\resizebox{\hsize}{!}{\includegraphics{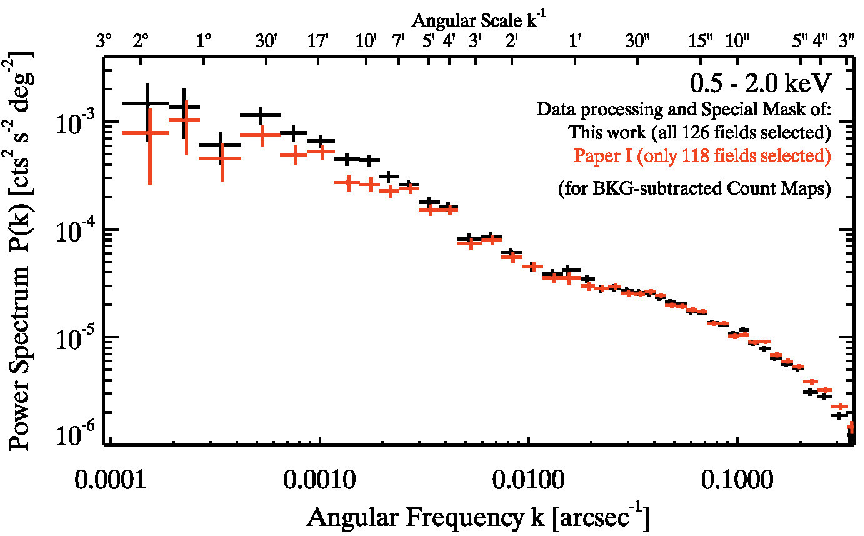}} 
				\caption{\label{f:PS_FieldSel}%
					Comparison of the combined \PoSpa{} in the \eSoftB{} for the data processing  of this work (black) and of \Kol{} (red) for the special mask.
					Note, that in this work all 126 observations of XBOOTES are considered,
					while for \Kol{} only 118 observations are considered (section~2 of \Kol{}).
					Also see \Sref{s:F_sel} and \Aref{a:Data} for details.
				}	
			\end{center}
			\end{figure}
			
			\begin{figure}
			\begin{center}
				\resizebox{\hsize}{!}{\includegraphics{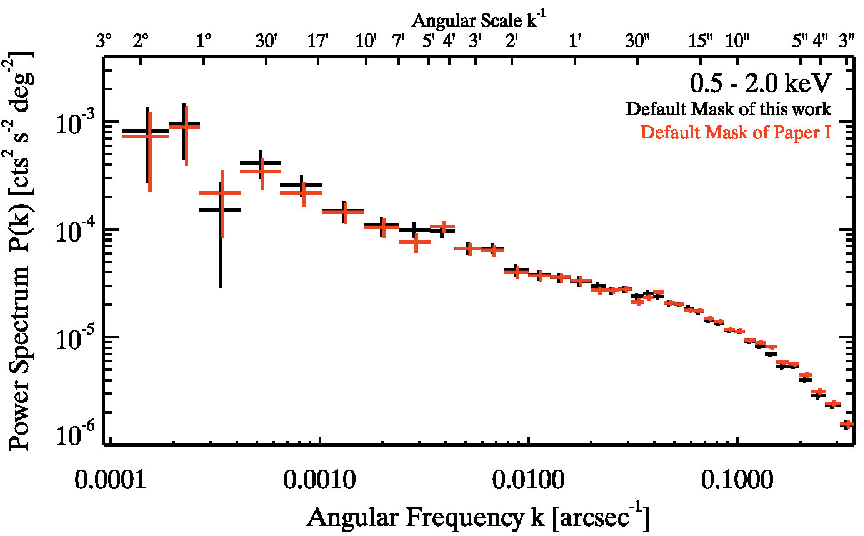}} 
				\caption{\label{f:PS_DefMsk}%
					Comparison of the combined \PoSpa{} in the \eSoftB{} for the default mask of this work (black) and of \Kol{} (red).
					The removal of \rPS{} is different (see \Sref{s:ResoPoSo} for details),
					while the removal of \rES{} is the same.
					Also see \Sref{s:ResoPoSo} and \Aref{a:Data} for details.
				}	
			\end{center}
			\end{figure}

		%-----------------------------------------------------------
		\subsection{Comparison of previous and current data processing} \label{a:Data}
		%-----------------------------------------------------------
		%\At{: Compute (eSoft, FOV92, OL00): DT02 with (RS01,RS10), also for 118 fields.}
		
		\Fref{PS_proc} shows that the stacked \PoSpa{} for the data processing of this work (black) and of \Kol{} (red) have overall a good agreement with each other for the default mask,
		although the field selection,
		exposure map and the FOV mask,
		the used count maps, 
		and the removal of \rPS{} have changed from \Kol{} to this work (\Sref{s:Change}).
		Note, that we rescaled the \PoSp{} of \Kol{} by the factor $(\langle E_\mathrm{Paper~I}\rangle/\langle E_\mathrm{This~work}\rangle)^2\approx0.86$,
		to account for the different average exposure times $\langle E\rangle$ of both data processing (\Sref{MSC:stat}),
		which allows us a better comparison of the shape of both \PoSpa{}.
		Given this good agreement for the default mask, we can concluded that also with the processed data of this work we would have come to the same result of \Kol{}.
		
		For the special mask the agreement is as good as for the default mask, if one keeps the field selection fixed to either the one from this work or from \Kol{} (\Sref{s:F_sel}).
		If one uses the default field selections, 118 observations for \Kol{} and all 126 observations for this work, there is however a significant disagreement of the corresponding \PoSpa{} on angular scales above $\gtrsim7\arcmin$.
		This can be seen in \Fref{PS_FieldSel}, where we show the combined \PoSpa{} to demonstrate that the field selection is also important on large angular scales above \acisi{}'s FOV ($\gtrsim17\arcmin$) for the special mask.
		Note, that in \Fref{PS_FieldSel} we had to use background-subtracted count maps ($\map{C}$) also for the \PoSpa{} of \Kol{} due the issues explained in \Aref{a:Bkg}.
		This discrepancy arises from the fact that the field selection of this work leads to a higher number of retained \rES{} than the field selection of \Kol{}, which was the main motivation to change the field selection. 
		This also means that the conclusions drawn from the \PoSp{} of the special mask in \Kol{} (section~5) would still remain the same with the current processed data.

		In this work we simplified the method of removing \rPS{} in comparison to \Kol{} (see \Sref{s:ResoPoSo} for details).
		\Fref{PS_DefMsk} shows that the combined \PoSpa{} of the default mask of this work (black) and of \Kol{} (red) agree very well with each other. 
		This agreement illustrates that both methods give consistent results on all considered angular scales.
		This is reassuring because with the simplified method of this work we reduce the size of the circular exclusion area, which leads to an increase in the number of residual counts per removed point source.
		If such an increase in residual counts would alter the \PoSp{} significantly then we would expect to see the largest differences on small angular scales.
		Fortunately, this is not the case for the given S/N.

	%===========================================================
	% Bibliography
	%===========================================================
			
\newcommand*\aap{A\&A}
\let\astap=\aap
\newcommand*\aapr{A\&A~Rev.}
\newcommand*\aaps{A\&AS}
\newcommand*\actaa{Acta Astron.}
\newcommand*\aj{AJ}
\newcommand*\ao{Appl.~Opt.}
\let\applopt\ao
\newcommand*\apj{ApJ}
\newcommand*\apjl{ApJ}
\let\apjlett\apjl
\newcommand*\apjs{ApJS}
\let\apjsupp\apjs
\newcommand*\aplett{Astrophys.~Lett.}
\newcommand*\apspr{Astrophys.~Space~Phys.~Res.}
\newcommand*\apss{Ap\&SS}
\newcommand*\araa{ARA\&A}
\newcommand*\azh{AZh}
\newcommand*\baas{BAAS}
\newcommand*\bac{Bull. astr. Inst. Czechosl.}
\newcommand*\bain{Bull.~Astron.~Inst.~Netherlands}
\newcommand*\caa{Chinese Astron. Astrophys.}
\newcommand*\cjaa{Chinese J. Astron. Astrophys.}
\newcommand*\fcp{Fund.~Cosmic~Phys.}
\newcommand*\gca{Geochim.~Cosmochim.~Acta}
\newcommand*\grl{Geophys.~Res.~Lett.}
\newcommand*\iaucirc{IAU~Circ.}
\newcommand*\icarus{Icarus}
\newcommand*\jcap{J. Cosmology Astropart. Phys.}
\newcommand*\jcp{J.~Chem.~Phys.}
\newcommand*\jgr{J.~Geophys.~Res.}
\newcommand*\jqsrt{J.~Quant.~Spec.~Radiat.~Transf.}
\newcommand*\jrasc{JRASC}
\newcommand*\memras{MmRAS}
\newcommand*\memsai{Mem.~Soc.~Astron.~Italiana}
\newcommand*\mnras{MNRAS}
\newcommand*\na{New A}
\newcommand*\nar{New A Rev.}
\newcommand*\nat{Nature}
\newcommand*\nphysa{Nucl.~Phys.~A}
\newcommand*\pasa{PASA}
\newcommand*\pasj{PASJ}
\newcommand*\pasp{PASP}
\newcommand*\physrep{Phys.~Rep.}
\newcommand*\physscr{Phys.~Scr}
\newcommand*\planss{Planet.~Space~Sci.}
\newcommand*\pra{Phys.~Rev.~A}
\newcommand*\prb{Phys.~Rev.~B}
\newcommand*\prc{Phys.~Rev.~C}
\newcommand*\prd{Phys.~Rev.~D}
\newcommand*\pre{Phys.~Rev.~E}
\newcommand*\prl{Phys.~Rev.~Lett.}
\newcommand*\procspie{Proc.~SPIE}
\newcommand*\qjras{QJRAS}
\newcommand*\rmxaa{Rev. Mexicana Astron. Astrofis.}
\newcommand*\skytel{S\&T}
\newcommand*\solphys{Sol.~Phys.}
\newcommand*\sovast{Soviet~Ast.}
\newcommand*\ssr{Space~Sci.~Rev.}
\newcommand*\zap{ZAp}		
	
	\footnotesize{
		\bibliographystyle{mn2e.bst} 
		%\bibliography{\LatDir Paper_References/references}
		% copy from paper_20160907.bbl
		
	}	
		
	%===========================================================
	% END of document
	%===========================================================
\label{lastpage}
\end{document}